\documentclass[12pt]{article}
\usepackage[utf8]{inputenc}

\usepackage{forest}
\usepackage[ruled,vlined]{algorithm2e}
\usepackage{mathtools}
\usepackage{graphicx}
\usepackage{multirow}
\usepackage{amssymb} 
\usepackage{amsthm}
\usepackage{amsmath}
\usepackage{fullpage}
\usepackage[longnamesfirst]{natbib}
\usepackage{enumerate}
\usepackage{amsfonts, bm, amsmath, color, natbib, rotating, amsthm, multicol, epstopdf, verbatim, hyperref}

\usepackage{authblk}
\usepackage{soul}
\bibpunct{(}{)}{;}{a}{,}{,}

\usepackage{setspace}
\usepackage{times}
    \usepackage{paralist}
    \usepackage[compact]{titlesec}
   \titlespacing{\section}{0pt}{0ex}{0ex}
    \titlespacing{\subsection}{0pt}{0ex}{0ex}
    \titlespacing{\subsubsection}{0pt}{0ex}{0ex}
    
\theoremstyle{plain} \newtheorem{theorem}{Theorem}  \newtheorem{lemma}{Lemma} \newtheorem{corollary}{Corollary}

\theoremstyle{definition}   \newtheorem{example}{Example}

\theoremstyle{remark}

\begin{document}

\def\T{{\mathrm{\scriptscriptstyle T} }}
\def\dsim{\mathrel{\dot\sim}}
\def\R{\mathbb{R}}
\def\C{\mathcal{C}}
\def\D{\mathcal{D}}
\def\E{\mathcal{E}}
\def\N{\mathbb{N}}
\def\Z{\mathcal{Z}}
\def\B{\mathbb{B}}
\def\pr{\mathbb{P}}
\def\Tau{\mathcal{T}}
\def\FF{\mathbf{F}}
\def\F{\mathcal{F}}
\def\M{\mathcal{M}}
\def\U{\mathcal{U}}
\def\X{\mathcal{X}}
\def\Y{\mathcal{Y}}
\def\SigmaE{\Sigma_\epsilon}
\def\iid{\stackrel{iid}{\sim}}

\newif\ifblinded

\ifblinded
\title{\Large Monte Carlo inference for semiparametric Bayesian regression}
\author{\vspace{-10mm}}
\else

\title{\Large Monte Carlo inference for semiparametric Bayesian regression}
\author{\large \vspace{-3mm} Daniel R. Kowal\thanks{Associate Professor, Department of Statistics and Data Science, Cornell University and  Department of Statistics, Rice University (\href{mailto:dan.kowal@cornell.edu}{dan.kowal@cornell.edu}).} \ and Bohan Wu\thanks{PhD student, Department of Statistics, Columbia University (\href{mail.to:bw2766@columbia.edu}{bw2766@columbia.edu})}} 

\fi

\date{}

\maketitle
  
  \vspace{-15mm}
\begin{abstract}
Data transformations are essential for broad applicability of parametric regression models. However, for Bayesian analysis, joint inference of the transformation and model parameters typically involves restrictive parametric transformations or nonparametric representations that are computationally inefficient and cumbersome for implementation and theoretical analysis, which limits their usability in practice. This paper introduces a simple, general, and efficient strategy for joint posterior inference of an unknown transformation and all regression model parameters. The proposed approach directly targets the posterior distribution of the transformation by linking it with the marginal distributions of the independent and dependent variables, and then deploys a Bayesian nonparametric model via the Bayesian bootstrap. Crucially, this approach delivers (1) joint posterior consistency under general conditions, including multiple model misspecifications, and (2) efficient Monte Carlo (not Markov chain Monte Carlo) inference for the transformation and all parameters for important special cases. These tools apply across a variety of data domains, including real-valued, positive, and compactly-supported data.  Simulation studies and an empirical application demonstrate the effectiveness and efficiency of this strategy for semiparametric Bayesian analysis with linear models, quantile regression, and Gaussian processes. 
\ifblinded 
The \texttt{R} package \texttt{[redacted]} is available on CRAN.
\else 
The \texttt{R} package \texttt{SeBR} is available on CRAN.
\fi 
\end{abstract}

\section{Introduction}
Transformations are widely useful for statistical modeling and data analysis. A well-chosen or learned transformation can significantly enhance the applicability of fundamental statistical modeling frameworks, such as Gaussian models \citep{box1964analysis}, linear and nonlinear regression models \citep{ramsay1988monotone,Carroll1988},  survival analysis \citep{Cheng1995}, discriminant analysis \citep{Lin2003}, and graphical models \citep{Liu2009}, among many other examples. This is especially true for data with complex marginal features (multimodality, skewness,  kurtosis, etc.) or data on different domains (real-valued, compactly-supported, or positive data), and for Bayesian probability models that must adapt to these attributes.

Consider the \emph{transformed regression model} for paired data  $\D_n = \{(x_i,y_i)\}_{i=1}^n$ with $x_i \in \mathcal{X} \subseteq \mathbb{R}^p$ and $y_i \in \mathcal{Y} \subseteq \mathbb{R}$:
\begin{align}
    \label{trans}
    g(y_i) &= z_i\\
    \label{mod}
    z_i &=  f_\theta(x_i) + \sigma \epsilon_i,\quad \epsilon_i \stackrel{iid}{\sim} P_\epsilon
\end{align}
where $g: \mathcal{Y}\to\mathbb{R}$ is a monotone increasing transformation, $f_\theta$ is a regression function with finite-dimensional parameters $\theta \in \Theta  \subseteq \R^d$, and $P_\epsilon$ is an  error distribution with fixed location and scale.  The regression model \eqref{mod}, denoted $P_{Z \mid \theta, x} \coloneqq f_\theta(x) + \sigma P_\epsilon$, may be considered the core statistical model, while the transformation  $g$ serves to improve the fidelity of this model for the given data $\D_n$. When the transformation is unknown, then \eqref{trans}--\eqref{mod} is a \emph{semiparametric regression model}. Identifiability is imposed on $P_{Z \mid \theta, x}$, typically by fixing the location ($f_\theta(0) = 0$) and scale ($\sigma = 1$).

We focus on Bayesian analysis  of \eqref{trans}--\eqref{mod}, but acknowledge the rich history of frequentist inference for transformation models, including monotone stress minimization \citep{Kruskal1965} 
alternating 
 conditional expectations \citep{Breiman1985},  additivity and variance stabilization \citep{Tibshirani1988}, transnormal regression models \citep{Fan2016}, and many others. 
 
There are several important examples of \eqref{trans}--\eqref{mod}:

\begin{example} \label{ex-gauss}
When $P_\epsilon =  N(0,1)$, model \eqref{mod} is a Gaussian regression:   $P_{Z \mid \theta, x} = N(f_\theta(x), \sigma^2)$. Although this model may be applied directly to $y$ and offers flexibility via $f_\theta$, the Gaussian assumption for the errors  is often restrictive and inadequate. The modeler must then consider whether to revise $f_\theta$, specify an alternative error distribution $P_\epsilon$, or incorporate a transformation via \eqref{trans}. We explore the latter option, and seek to provide excellent empirical performance, efficient algorithms, and strong theoretical guarantees for Bayesian inference. 
\end{example}

\begin{example} \label{ex-quantile}
Bayesian quantile regression specifies $P_\epsilon$ such that $f_\theta(x)$ target the $\tau$th quantile of $P_{Z \mid \theta, x}$. The most common choice is the asymmetric Laplace distribution (ALD) with density $p_\tau(\epsilon) = \tau(1-\tau)\exp\{-\rho_\tau(\epsilon)\}$ and $\rho_\tau(\epsilon) = \epsilon\{\tau - I(\epsilon <0)\}$ is the check loss function  \citep{Yu2001}. However, the ALD is often a poor model for data, especially when $\tau$ is close to zero or one (see Section~\ref{sims-qr}). 
A transformation can alleviate such inadequacy. The $\tau$th quantile of $z$ corresponds to the $g^{-1}(\tau)$th quantile of $y$, so  $f_\theta(x)$ maintains interpretability and the transformed regression model readily provides quantile estimates for $y$ at $x$.
\end{example} 


The transformation $g$ plays two critical roles. First, it increases modeling flexibility to  better satisfy common assumptions for $P_{Z\mid \theta, x}$, such as linearity, additivity, homoscedasticity, or Gaussianity \citep{box1964analysis,Lin2020},  improve prediction \citep{DeOliveira1997}, and  handle various marginal distributions of $y$, including those with multimodality, skewness,  kurtosis, etc. Second, $g$ delivers probabilistic coherency for the support $\mathcal{Y}$,   such as real-valued data $\Y = \mathbb{R}$, compactly-supported data $\Y = [0,1]$, or positive data $\Y = \mathbb{R}^+$. In the examples above, the model $P_{Z \mid \theta, x}$ is supported  on $\mathbb{R}$. Thus, a transformation admits broader  applicability to other data domains.
Here, we assume that $\mathcal{Y}$ is continuous; the integer case of $\Y = \mathbb{Z}$ is addressed in  \cite{Kowal2021c}, but only for Gaussian linear models for \eqref{mod}. 
Our analysis is significantly broader and more robust, and includes new methods and theory for general settings plus tailored algorithms and detailed analysis for linear variable selection, quantile regression, and 
Gaussian processes.



For Bayesian analysis, an unknown transformation must be modeled and accounted for with the joint posterior distribution $p(g, \theta  \mid \D_n)$. 
Parametric specifications of $g$ such as the extended Box-Cox family \citep{Atkinson2021} are widely popular, especially in conjunction with regression or Gaussian process models for \eqref{mod} \citep{Pericchi1981,Lin2020}. However, parametric transformations sacrifice modeling flexibility and may not be suitable for some data domains. Furthermore, parametric transformations do not guarantee computational convenience for the joint posterior  $p(g, \theta  \mid \D_n)$:  in fact, $g$ is typically  fixed at a point estimate \citep{DeOliveira1997,bean2016transformations,rios2018learning,Lin2020}. This practice ignores the uncertainty in the transformation. An exception is \cite{gottardo2009bayesian}, who used a  Metropolis-within-Gibbs Markov chain Monte Carlo (MCMC) algorithm. 





    

Nonparametric models for $g$ include 
Gaussian processes \citep{lazaro2012bayesian}, mixtures of incomplete beta or hyperbolic functions \citep{mallick1994generalized,Mallick2003,Snelson2003},  splines 
\citep{Wang2011a,Song2012,Tang2018,Wu2019a,Mulgrave2018,Kowal2020a}, normalizing flows \citep{Maronas2021}, or compositions \citep{Rios2019}. Each of these models requires constraints to ensure monotonicity of $g$. 
More critically, these  approaches do not provide easy access to the joint posterior $p(g, \theta \mid \D_n)$. A common MCMC strategy is to use Metropolis-within-Gibbs \citep{mallick1994generalized,Mallick2003,Song2012,Kowal2020a} or Hamiltonian Monte Carlo (HMC)-within-Gibbs \citep{Mulgrave2018} to iteratively sample  $[g \mid \D_n, \theta]$ and $[\theta \mid \D_n, g]$. However, the sampling steps for $[g \mid \D_n, \theta]$ are typically complex and require careful tuning to ensure adequate MCMC performance. 
Alternatively, a full Gibbs sampler cycles individually through each parameter that determines $g$, sampling each from its univariate full conditional distribution \citep{Wang2011a,Tang2018,Wu2019a}. However, Gibbs samplers with such small parameter blocks sacrifice Monte Carlo efficiency relative to joint samplers.   
Finally, variational approximations \citep{lazaro2012bayesian,Maronas2021}
achieve computational efficiency but are often inadequate for uncertainty quantification \citep{blei2017variational}. 
These factors limit the utility of existing approaches for Bayesian analysis of  \eqref{trans}--\eqref{mod}. 









    




The proposed approach bears some resemblance to  copula-based methods that decouple marginal and joint parameter estimation, called inference function for margins \citep{Joe2005}. That framework uses a two-stage point estimation that is inadequate for joint uncertainty quantification, and predominantly is limited to copula models. \cite{Grazian2017} introduced a Bayesian analog, which uses an empirical likelihood approximation with MCMC. \cite{Klein2019} and \cite{Smith2021} applied copula models for regression analysis, but again relied on  MCMC  or variational approximations. Alternatively, the rank likelihood \citep{Pettitt1982} eschews estimation of $g$ and provides inference for $\theta$ based only on the ranks of $y$. However, this approach does not produce a coherent posterior predictive distribution,  requires computationally demanding MCMC sampling,  and primarily focuses on copula models  \citep{Hoff2007,Feldman2021}.

Conditional transformation models (CTMs) have firmly established the utility of transformation models for distributional regression analysis \citep{hothorn2014conditional}. In comparison with the proposed approach, CTMs feature a conditional (on $X$) rather than marginal transformation $g$, and anchor the model on a marginal rather than conditional distribution $P_{Z \mid\theta, x}$. By design,  \eqref{trans}--\eqref{mod} aligns more closely with, and suitably formalizes, the ``transform, then model" approach that is widely adopted in practice. 
Although CTMs offer substantial flexibility for modeling $Y \mid X$, Bayesian CTMs \citep{carlan2023bayesian} feature nonstandards likelihoods and tensor product basis expansions with monotonicity constraints. Thus, posterior inference is challenging, and relies on complex and carefully-tuned HMC/MCMC sampling algorithms.

This manuscript introduces a general methodological and computational framework for Bayesian inference and prediction for the transformed regression model \eqref{trans}--\eqref{mod}. The proposed approach is easy to implement for a variety of useful regression models and delivers efficient Monte Carlo (not MCMC) inference (Section~\ref{sec-methods}). Empirically, this framework improves prediction, variable selection, and estimation of $g$ for Bayesian semiparametric linear models; provides substantially more accurate quantile estimates and model adequacy for Bayesian quantile regression; and increases predictive accuracy for Gaussian processes (Section~\ref{sec-emp}). Our theoretical analysis establishes and characterizes posterior consistency, including multiple model misspecifications 
(Section~\ref{sec-theory}). Some limitations are addressed in Section~\ref{sec-disc}. 
Supplementary material includes technical proofs, additional simulation results, and reproducible \texttt{R} code. 
\ifblinded 
An \texttt{R} package \texttt{[redacted]} is on CRAN with detailed documentation and examples at  \texttt{[redacted]}.
\else 
An \texttt{R} package \texttt{SeBR} is on CRAN with detailed documentation and examples at \url{https://drkowal.github.io/SeBR/}.
\fi

 
%





    


\section{Methods} \vspace{-1mm}  \label{sec-methods}
\subsection{General case}\label{sec-general}
The goal is to provide Bayesian inference for transformed regression models \eqref{trans}--\eqref{mod}, prioritizing (i) nonparametric modelling of the transformation $g$, (ii) computational convenience and efficiency, and (iii) empirical and theoretical validation for  posterior and predictive inference. 
Our approach begins with a general decomposition of the joint posterior distribution:
\begin{equation}
    \label{decomp}
    p(g, \theta \mid \D_n) =  p(g  \mid \D_n) \ p(\theta  \mid \D_n, g).
\end{equation}
The second term is straightforward: $p(\theta  \mid \D_n, g)$ is equivalent to the posterior distribution of $\theta$ under model \eqref{mod} using data $z_i = g(y_i)$  with known transformation $g$. The presence of the transformation does not introduce any additional challenges for this term: the conditional posterior $p(\theta \mid \D_n, g) = p(\theta  \mid \{x_i, g(y_i)\}_{i=1}^n)$ is well-studied for many regression models \eqref{mod} and priors $p(\theta)$, and is either available analytically or estimable using standard algorithms for Bayesian regression.  The first term,   $p(g \mid \D_n)$, presents the more significant challenge. 
A direct Bayesian approach---specifying a prior for $(g, \theta)$, computing the likelihood under \eqref{trans}--\eqref{mod}, and attempting to marginalize the posterior,  $p(g \mid \D_n) = \int_\Theta p(g, \theta \mid \D_n) \ d\theta$---is analytically intractable, while MCMC-based approximations are complex, computationally inefficient, and empirically unsatisfactory (see Section~\ref{sec-emp}). Thus, we pursue alternative strategies to access or approximate $p(g \mid \D_n)$.



The central idea is that, under model \eqref{trans}--\eqref{mod},  we can infer $g$ indirectly by learning the \emph{marginal} distributions $P_Y$ and $P_X$  of the dependent and independent variables, respectively---and that this approach is especially fruitful for Bayesian computing, inference, and theory. 
Under \eqref{trans}, the  links among $g$,  $P_Y$, and $P_X$  are established 
via  the cumulative distribution functions (CDFs):   $    F_{Y}(t)  = \mbox{pr}(y \le t) =  \mbox{pr}\{z \le g(t)\} = F_{Z}\{g(t)\}$, so the transformation is 
\begin{equation}
    \label{trans-cdf}
    g(t) = F_{Z}^{-1}\{F_{Y }(t)\}.
\end{equation}
These CDFs must be defined carefully in accordance with model \eqref{mod}. Each term in \eqref{trans-cdf} is invariant to $\theta$ and $X$. Thus, while $F_Y$ is the familiar marginal CDF of $y$, $F_Z$ is \emph{not} simply the regression CDF $F_{Z \mid \theta, x}$ from \eqref{mod}. Instead, marginalization over  $\theta$ and $X$ is required:   $F_Z(t) \coloneqq   \mathbb{E}_{\theta, X}\{F_{Z \mid \theta, X}(t)\}
$. For illustration, consider \emph{a priori} analysis, for example, to construct a nonparametric prior for $g$.  Suppose the covariates are fixed, so $P_X$ is the empirical distribution of $\{x_i\}_{i=1}^n$. Then $F_Z(t) = n^{-1} \sum_{i=1}^n \mathbb{E}_{\theta}\{F_{Z \mid \theta, X = x_i}(t)\}$, where the summand is the marginal (or prior predictive) CDF of the regression model \eqref{mod} with prior $\theta \sim p(\theta)$. Thus, a nonparametric prior for   $P_Y$ (e.g., a Dirichlet process prior) is sufficient to induce a nonparametric prior for $g$ via \eqref{trans-cdf}. This prior is also generative: simply draw $P_Y$ from its prior and compute  \eqref{trans-cdf} using the implied CDF $F_Y$ with $F_Z$ as defined above. Notably, $g$ is monotone by construction, with no further constraints needed.



We adapt and extend these ideas for \emph{posterior} inference for $g$. Informally, we propose a generative algorithm to (approximately) sample from $p(g \mid \D_n)$ via  \eqref{trans-cdf}, where each constituent term is drawn from a suitable posterior distribution. The resulting posterior and predictive distributions are then validated empirically (Section~\ref{sec-emp}) and theoretically (Section~\ref{sec-theory}) for a variety of regression models \eqref{mod}. To ensure nonparametric flexibility for $g$, we use Bayesian nonparametric models for the required distributions: $P_Y$ and, if the covariates are random, $P_X$. Among the many appealing options \citep{Ghosal2017}, we prioritize models that admit efficient computing and theoretical analysis for $g$, and thus must carefully consider the implications for $F_Y$ and $F_Z$ in \eqref{trans-cdf}.

First, consider $P_Y$. Our default recommendation is the Bayesian bootstrap (BB), which may be constructed by placing a Dirichlet process prior over $\mathcal{Y}$ and taking the limit as the concentration parameter goes to zero \citep{rubin1981bayesian}. The BB  requires no tuning parameters, applies for various data domains, and offers substantial modeling flexibility---all while admitting an efficient posterior sampling algorithm (Algorithm~\ref{alg:Fy}). Notably,  Algorithm~\ref{alg:Fy} features Monte Carlo rather than MCMC sampling, and thus avoids the need for lengthy runs and convergence diagnostics while still controlling the approximation error via the number of simulations. 
\begin{algorithm}
\begin{enumerate}

    \item Sample $(\alpha_1^y,\ldots,\alpha_n^y) \sim \mbox{Dirichlet}(1,\ldots,1)$
    \vspace{-2mm}
    
    \item Compute $\tilde F_{Y \mid \D_n}(t) =   \sum_{i=1}^n \alpha_i^y  I(y_i \le t)$
    \vspace{-2mm}
    
\end{enumerate} 
\caption{Monte Carlo sampling of $F_{Y \mid \D_n}$ for the Bayesian bootstrap.
 \label{alg:Fy}}
\end{algorithm}  

Next, consider $P_X$. Usually, Bayesian regression models  circumvent the need to specify $P_X$  by assuming independence between $\theta$ and the parameters that govern $P_X$ \citep{Gelman1995}. Here, $P_X$ is required only for the marginalization that determines $F_Z$ in \eqref{trans-cdf}. Specifically, $F_Z$ again requires marginalization over $X$ and $\theta$, but now using the respective posteriors: 
\begin{equation}
    \label{z-cdf-1}
    F_{Z\mid \D_n}(t) \coloneqq    \mathbb{E}_{\theta, X \mid \D_n}\{F_{Z \mid \theta, X}(t)\}
    =  \int_\mathcal{X}  F_{Z \mid X=x}(t)  \ d P_{X}(x) 
\end{equation}
where $F_{Z \mid \theta, X}$ is the regression CDF from \eqref{mod}, $F_{Z \mid X}(t) \coloneqq \int_\Theta  F_{Z \mid \theta, X}(t) \ p(\theta \mid \D_n) \ d \theta$, and we have assumed independence between $\theta$ and $X$  (i.e., $\theta$ does not appear in the model for $P_X$). For $P_X$, we recommend another BB: it is automated, flexible, applies to mixed data types, and delivers an exceptionally convenient and efficient Monte Carlo (not MCMC) sampling algorithm  for   $F_{Z\mid\D_n}$  (Algorithm~\ref{alg:Fz}). In particular, with Algorithm~\ref{alg:Fz}, the required integration over $X$ in  \eqref{z-cdf-1}---while accounting for posterior uncertainty about $P_X$---reduces to a simple summation. 
When the covariates are fixed, we use the empirical distribution for $P_X$ ($\alpha_i^x = 1/n$) and thus $F_{Z \mid \D_n}$ is deterministic (note that $z$ in \eqref{mod} is still random with $\sigma \ne 0$). 
\begin{algorithm} 
\begin{enumerate}
    
    \item Sample $(\alpha_1^x,\ldots,\alpha_n^x) \sim \mbox{Dirichlet}(1,\ldots,1)$
    \vspace{-2mm} 
    
    \item Compute $\tilde F_{Z \mid \D_n}(t) = \sum_{i=1}^n \alpha_i^x F_{Z \mid X=x_i}(t)$
    \vspace{-2mm}
    
\end{enumerate} 
\caption{Monte Carlo sampling of $F_{Z \mid \D_n}$ for the Bayesian bootstrap. \label{alg:Fz}} 
\end{algorithm} 



By combining draws of $\tilde F_{Y\mid \D_n}$ (Algorithm~\ref{alg:Fy}) and $\tilde F_{Z \mid \D_n}$ (Algorithm~\ref{alg:Fz}) and computing \eqref{trans-cdf}, we obtain a simple, general, and efficient Monte Carlo (not MCMC) algorithm that targets $p(g \mid \D_n)$. This approach is customizable for different regression models \eqref{mod} via  $F_{Z\mid X}$ (Sections~\ref{sec-sblm}--\ref{sec-sbgp}). Although this term  also depends on $p(\theta \mid \D_n)$, we provide effective approximation strategies and crucially show that the proposed sampler is robust to these approximations  
(Section~\ref{sec-approx}). For instance, even using the \emph{prior} $p(\theta)$ for this approximation  produces highly competitive empirical results (Section~\ref{sec-emp}). We confirm this robustness asymptotically  in Section~\ref{sec-theory}.

We present our main algorithm for joint posterior  and predictive inference in Algorithm~\ref{alg:joint}. Here, the joint posterior decomposition \eqref{decomp} is augmented with a posterior predictive variable $\tilde y(x)$, i.e., the posterior distribution of future or unobserved data at $x$ according to the model \eqref{trans}--\eqref{mod}:   $ p\{g, \theta, \tilde y(x) \mid \D_n\} =  p(g  \mid \D_n) \ p(\theta  \mid \D_n, g)\ p\{\tilde y(x)  \mid \D_n, g, \theta\}.$
\begin{algorithm}
\begin{enumerate}
    \item Simulate $g^* \sim p(g  \mid \D_n)$:
    \vspace{-2mm}
    
    \begin{enumerate}
        \item Sample $\tilde F_{Z\mid\D_n}$ (Algorithm~\ref{alg:Fz})
        \vspace{-1mm}
        
        \item Sample $\tilde F_{Y\mid\D_n}$ (Algorithm~\ref{alg:Fy})
        \vspace{-1mm}
        
        \item Compute $ g^*(t) = \tilde F_{Z\mid\D_n}^{-1}\{ n(n+1)^{-1}\tilde F_{Y\mid\D_n}(t)\} $
        \vspace{-2mm}
    \end{enumerate}

    \item Simulate $\theta^* \sim  p(\theta \mid \D_n, g = g^*)$
    \vspace{-2mm}

    \item Simulate $\tilde y^*(x) \sim p\{\tilde y(x) \mid \D_n, g = g^*, \theta = \theta^*\}$:
    \vspace{-2mm}
    
    \begin{enumerate}
        \item Sample $\epsilon^* \sim P_\epsilon$ and compute 
         $\tilde z^*(x) = f_{\theta^*}(x) + \sigma \epsilon^*$
        \vspace{-1mm}
        
        \item Compute $\tilde y^*(x) = (g^*)^{-1}\{\tilde z^*(x)\}$
        \vspace{-2mm}
        
    \end{enumerate}
    
\end{enumerate} 
\caption{Joint Monte Carlo sampler for $(g^*, \theta^*, \tilde y^*(x)) \sim p\{g, \theta, \tilde y(x) \mid \D_n\}$.
 \label{alg:joint}}
\end{algorithm} 
Algorithm~\ref{alg:joint} is a Monte Carlo sampler for $(g, \theta, \tilde y(x))$ jointly whenever the sampler for $p(\theta \mid \D_n, g = g^*)$---equivalently, the posterior distribution from model \eqref{mod} using data $z_i^* = g^*(y_i)$---is Monte Carlo. 
Even when approximate sampling algorithms are required for  the posterior of $\theta$, Algorithm~\ref{alg:joint} crucially avoids a Gibbs blocking structure for $[g \mid \D_n, \theta]$ and $[\theta \mid \D_n, g]$, which distinguishes the proposed approach from existing sampling algorithms. 
The monotonicity of each sampled $g^*$ is guaranteed by construction.

The role of $n(n+1)^{-1}$  is to avoid boundary issues: when the latent data model \eqref{mod} is supported on $\R$, 
$\tilde F_{Z\mid\D_n}^{-1}(1) = \infty$ for any $t$ such  that    $\tilde F_{Y\mid \D_n}(t) = 1$. Under the BB for $P_Y$, this occurs for $t \ge \max\{y_i\}$ and thus cannot be ignored. The rescaling eliminates this nuisance to ensure finite $g$ but is asymptotically negligible. 

The predictive sampling step requires application of  $(g^*)^{-1}(s) = \tilde F_{Y\mid\D_n}^{-1} \{\tilde F_{Z\mid\D_n}(s)\}$. Thus, the posterior predictive distribution matches the support of $P_Y$. However, the BB for $P_Y$ is supported only on the observed data values $\{y_i\}_{i=1}^n$, even though $\mathcal{Y}$ is continuous. We apply a monotone and smooth interpolation of $g^*$ \citep{Fritsch1980} prior to computing its inverse, which only impacts the predictive sampling step---not the sampling of $\theta$---yet expands the support of $\tilde y(x)$ to $[\min(y), \max(y)]$. These endpoints may be extended as appropriate.

\subsection{Correction factors and robustness adjustments}\label{sec-approx}
Algorithm~\ref{alg:joint} faces two noteworthy obstacles. First, the sampler for $p(g \mid \D_n)$ does not use the exact likelihood under model \eqref{trans}--\eqref{mod}. We characterize this discrepancy with the following result. 
\begin{theorem}\label{thm-adj}
Suppose that $F_Y$ and $F_Z$ are continuous distribution functions with densities that exist. Under  model \eqref{trans}--\eqref{mod} with $\theta \sim p(\theta)$ and $x_i \sim P_X$ independently, the likelihood for $g$ is
\begin{equation}\label{exact-like}
p(y_1,\ldots, y_n \mid g) =  
\frac{
\mathbb{E}_\theta  [\prod_{i=1}^n p_{Z \mid \theta}\{g(y_i)\} ]
}{
\prod_{i=1}^n \mathbb{E}_\theta[p_{Z \mid \theta}\{g(y_i)\}]
}\prod_{i=1}^n p_Y(y_i)
= \omega(y \mid g)\prod_{i=1}^n p_Y(y_i)
\end{equation}
where $\mathbb{E}_\theta$ is the expectation under $p(\theta)$ and $\omega(y \mid g)$ is a  correction factor.
\end{theorem}
The correction factor $\omega(y \mid g)$ appears because model \eqref{trans}--\eqref{mod} implies (marginal) exchangeability but not independence for $\{y_i\}_{i=1}^n$. Algorithm~\ref{alg:Fy} intentionally omits $\omega(y \mid g)$ and instead uses only $\prod_{i=1}^n p_Y(y_i)$, which we refer to as the \emph{surrogate} likelihood for $g$. The remaining sampling steps use the correct likelihood. 
This strategy is fruitful: it delivers efficient Monte Carlo (not MCMC) sampling (Algorithm~\ref{alg:joint}) and consistent posterior inference for $g$ (Section~\ref{sec-theory}), which further indicates that the omitted term $\omega(y \mid g)$ is indeed asymptotically negligible. 


It is possible to correct for the surrogate likelihood using importance sampling.  First,  we apply Algorithm~\ref{alg:joint} to obtain $S$ draws $\{g^s, \theta^s, \tilde y^s(x)\}_{s=1}^{S} \sim p\{g, \theta, \tilde y(x) \mid \D_n\}$. Using this as the proposal distribution, the importance weights are  $\omega(y \mid g)$,  and   may be used to estimate expectations or obtain corrected samples via sampling importance resampling. The latter version draws indices $\{s_1^*, \ldots, s_{S^*}^*\} \subset \{1,\ldots, S\}$ with replacement proportional to  $\omega(y \mid g^s)$ and retains the subsampled draws $\{g^s, \theta^s, \tilde y^s(x)\}_{{s}=s_1^*}^{s_{S^*}^*}$ with $S^* < S$. 
However, our empirical results (see the supplementary material)  suggest that even for $n=50$, this adjustment has minimal impact and is not necessary to achieve excellent  performance.

The second challenge pertains to $F_{Z \mid \D_n}$ in \eqref{z-cdf-1} and Algorithm~\ref{alg:Fz}, which depends on $p(\theta \mid \D_n)$. At first glance, this is disconcerting: the posterior of $\theta$ under model \eqref{trans}--\eqref{mod}---unconditional on the transformation $g$---is not easily accessible. However, a critical feature of the proposed approach (Algorithm~\ref{alg:joint}) is that posterior and predictive inference is remarkably robust to approximations of $p(\theta \mid \D_n)$ in \eqref{z-cdf-1}. 
In particular, this quantity is merely one component that defines $g$ in \eqref{trans-cdf}, and the  remaining  posterior and predictive sampling steps in Algorithm~\ref{alg:joint} use the exact (conditional) posterior $p\{\theta, \tilde y(x) \mid \D_n, g\}$. In fact, we show empirically that using the \emph{prior}  $p(\theta)$ as a substitute for $p(\theta \mid \D_n)$ in  \eqref{z-cdf-1} yields highly competitive results, even with noninformative priors (Section~\ref{sec-emp}). 

The general idea is to substitute an approximation $\hat p(\theta  \mid \D_n)$ into Algorithm~\ref{alg:Fz} via 
\begin{equation}\label{Fzx}
    \hat F_{Z \mid X}(t) = \int_\Theta  F_{Z \mid \theta, X}(t) \ \hat p(\theta \mid \D_n) \ d \theta
\end{equation}
which modifies \emph{only} the sampling step for $p(g \mid \D_n)$ in Algorithm~\ref{alg:joint} and not the subsequent draws of $\theta$ or $\tilde y(x)$. When \eqref{Fzx} is not available analytically, we may estimate it using Monte Carlo:  $\hat F_{Z \mid X}(t) \approx S^{-1} \sum_{s=1}^S F_{Z \mid \theta = \theta^s, X}(t) $ where $\{\theta^s\}_{s=1}^S \sim \hat p(\theta \mid \D_n)$ and $F_{Z \mid \theta, X}$ corresponds to \eqref{mod}.

We consider three options for $\hat p(\theta \mid\D_n)$: (i) the prior $p(\theta)$, (ii) rank-based procedures that directly target $p(\theta \mid \D_n)$ without estimating  $g$ \citep{Horowitz2012}, and (iii) plug-in approximations that target $p(\theta \mid \D_n, g = \hat g)$ for some point estimate $\hat g$. The rank-based approaches offer appealing theoretical properties, but are  significantly slower, designed primarily for linear models,  and produce nearly identical initial estimates and results as our implementation of (iii) (see the supplementary material). Thus, we focus on (i) and (iii)  and discuss (ii) in the supplementary material. 


Option (iii)  considers approximations of  the form $p(\theta \mid \D_n, g = \hat g) = p(\theta \mid \{x_i, \hat g(y_i)\}_{i=1}^n)$, which is the  posterior under model \eqref{mod} given data $\{x_i, \hat g(y_i)\}_{i=1}^n$ and a fixed transformation, such as
  $  \hat g(t) = \hat F_{Z}^{-1}\{n(n+1)^{-1} \hat F_{Y}(t)\}.$ 
Many approximation strategies  exist for  $p(\theta \mid \D_n, g = \hat g)$; our default is the fast and simple  Laplace approximation $\hat p(\theta \mid \D_n, g = \hat g) = N(\hat \theta, \sigma^2 \Sigma_{\hat\theta})$, where $\hat \theta$ is the posterior mode and $\sigma^2\Sigma_{\hat\theta}$ approximates the posterior covariance using data $\{x_i, \hat g(y_i)\}_{i=1}^n$. To compute $\hat g$, we let $\hat F_{Y}(t) = n^{-1}  \sum_{i=1}^n   I(y_i \le t)$  be the empirical CDF of $y$ and update $\hat F_{Z}$ in two stages. First, we  initialize $\hat g_0(t) = \Phi\{n(n+1)^{-1}\hat F_Y(t)\}$ 
and then compute $\hat p(\theta \mid \D_n, g = \hat g_0)$. From this initialization, we update $\hat F_{Z}(t) = n^{-1} \sum_{i=1}^n  \hat F_{Z \mid X=x_i}(t)$ using $\hat F_{Z \mid X}$ in \eqref{Fzx}, which supplies an updated transformation $\hat g$ and thus an updated  approximation $\hat p(\theta \mid \D_n, g = \hat g)$. Finally, this approximation is use for \eqref{Fzx} and substituted into Algorithm~\ref{alg:Fz}, and is a one-time cost for all samples  of $p(g \mid \D_n)$ in Algorithm~\ref{alg:joint}. Again, this approximation utilized only for  sampling $p(g \mid \D_n)$, while the remaining steps of Algorithm~\ref{alg:joint} use the exact (conditional) posterior $p\{\theta, \tilde y(x) \mid \D_n, g\}$.

Lastly, to accompany and partially correct for the approximation $\hat p$ in \eqref{Fzx}, we add a simple yet effective robustness adjustment to Algorithm~\ref{alg:joint}. First, observe that the location and scale  of the latent data model \eqref{mod} map to the location and scale of the transformation, and vice versa:  
\begin{lemma}\label{g-scale}
Consider a transformation $g_{1}$ that uses the  distribution of $[\mu + \sigma Z \mid \D_n]$ instead of 
$[Z \mid \D_n]$, where $\mu$ and $\sigma$ are fixed constants. Then $g_1(t) = \mu + \sigma g(t)$.
\end{lemma}
Lemma~\ref{g-scale} is not merely about identifiability, but also suggests a triangulation strategy for robustness to  $\hat p$. If  $\hat p$  induces the wrong location or scale for $F_{Z \mid \D_n}$ via \eqref{z-cdf-1} and \eqref{Fzx}, then this misspecification propagates to $p(g \mid \D_n)$  and then $p\{\theta, \tilde y(x) \mid \D_n, g\}$. 
The key insight from Lemma~\ref{g-scale} is that this effect is reversible: the wrong location or scale for $p(g \mid \D_n)$---regardless of the source---can be \emph{corrected} by suitably adjusting  the location and scale in the sampling step for $p\{\theta, \tilde y(x)\mid \D_n, g\}$. We modify 
Algorithm~\ref{alg:joint} accordingly. Step~1 is unchanged: we sample from  $p(g \mid \D_n)$ under an identified model with fixed location ($f_\theta(0) = 0$) and scale ($\sigma =1$) in \eqref{mod}. For Steps~2~and~3, we reintroduce the location and scale in \eqref{mod} by replacing $P_{Z \mid \theta, x}$ with $\mu + \sigma P_{Z \mid \theta, x}$, where $(\mu, \sigma)$ is assigned a diffuse prior, and then sample from the joint posterior (predictive) distribution   $p\{\mu, \sigma,\theta, \tilde y(x) \mid \D_n, g\}$. This modification is simple and convenient---regression models \eqref{mod} typically include location and scale parameters---yet provides inference for $(\theta, \tilde y(x))$ that is robust against location-scale misspecification of $p(g \mid \D_n)$.
 We confirm this effect both empirically (Section~\ref{sec-emp}) and theoretically (Section~\ref{sec-theory}).


These location-scale parameters $(\mu, \sigma)$ are not identified under model \eqref{trans}--\eqref{mod}. Thus, they are not strictly necessary in any analysis or in Algorithm~\ref{alg:joint}. We  do not seek to interpret them. Instead, we view $(\mu, \sigma)$ as an accompaniment to the approximation $\hat p$, which determines $p(g \mid \D_n)$ via \eqref{Fzx} and Algorithm~\ref{alg:Fz}. Instead of supplying a more sophisticated approximation $\hat p$ to infer $g$, Lemma~\ref{g-scale} suggests that we may equivalently apply a downstream location-scale adjustment in the regression model  \eqref{mod}. 
This motivates our use of $\mu + \sigma P_{Z \mid \theta, x}$.

\subsection{Semiparametric Bayesian linear regression}\label{sec-sblm}
Suppose that \eqref{mod} is the Gaussian linear regression model $P_{Z \mid \theta, X=x} = N(x^\T \theta, \sigma^2)$ with prior $\theta \sim N(\mu_\theta, \sigma^2\Sigma_\theta)$. We focus on the  $g$-prior \citep{Zellner1986} with $\mu_\theta = 0$ and $\Sigma_\theta = \psi  (X^\T X)^{-1}$ for $\psi > 0$. 
For model identifiability, the scale is fixed at $\sigma=1$ and the intercept is omitted.  
We  apply Algorithm~\ref{alg:joint} to obtain Monte Carlo posterior draws of $(g^*, \theta^*, \tilde y^*(x))$ as follows.

First, consider $g$. The necessary step is to construct $\hat F_{Z\mid X = x_i}$  and apply Algorithm~\ref{alg:Fz} to sample $g^* \sim p(g \mid \D_n)$. Using a preliminary approximation of the form $\hat p(\theta \mid \D_n, g = \hat g) = N(\hat \theta,  \Sigma_{\hat\theta})$, this key term is $\hat F_{Z \mid X = x_i}(t) = \Phi(t; x_i^\T\hat\theta, 1 + x_i^\T  \Sigma_{\hat\theta} x_i)$. We consider two options for $\hat p$: the prior, $\hat \theta = 0$ and $\Sigma_{\hat\theta} = \psi (X^\T X)^{-1}$, or the Laplace approximation, $ \hat \theta = 
\Sigma_{\hat\theta}X^{\T} \hat g(y)$ with $\Sigma_{\hat\theta} = 
\psi(1+\psi)^{-1}(X^\T X)^{-1}$. 
The remaining steps for sampling $g^* \sim p(g \mid \D_n)$ are straightforward. 

Next, we sample $p(\theta \mid \D_n, g = g^*)$. Including the robustness adjustment, we add an intercept parameter to $\theta$ and assign a prior to the scale, $\sigma^{-2} \sim \mbox{Gamma}(a_\sigma, b_\sigma)$, with small values for the hyperparameters ($a_\sigma = b_\sigma = 0.001$). We \emph{jointly} sample $p(\sigma, \theta \mid \D_n, g ) = p(\sigma \mid \D_n, g)\ p(\theta  \mid \D_n, g, \sigma)$, where $[\sigma^{-2} \mid \D_n, g = g^*] \sim \mbox{Gamma}(a_\sigma + n/2, b_\sigma + \{\Vert z^* \Vert^2 -  \psi(1+\psi)^{-1} (z^*)^\T X(X^\T X)^{-1} X^\T z^*\}/2)$ for $z_i^* = g^*(y_i)$ and $[\theta \mid \D_n, g = g^*, \sigma = \sigma^*] \sim N(Q_\theta^{-1} \ell_\theta, Q_\theta^{-1})$ with $Q_\theta = (\sigma^*)^{-2}(1+\psi)\psi^{-1} X^\T X $ and $\ell_\theta = (\sigma^*)^{-2} X^\T z^*$. Even with the location-scale adjustment, all quantities are drawn jointly using Monte Carlo (not MCMC) sampling.

Finally, the predictive sampling step is  $\tilde z^*(x) \sim N(x^\T \theta^*, (\sigma^*)^2)$ and $\tilde y^*(x) = (g^*)^{-1}\{\tilde z^*(x)\}$.





To apply the importance sampling adjustment \eqref{exact-like}, we may use the sampled weights $\alpha^x$ to target the  densities that determine $\omega(y \mid g)$:
\begin{equation}\label{imp-linear}
    \omega(y \mid g) \approx \frac{
    S^{-1}\sum_{s=1}^S \prod_{i=1}^n \sum_{i'=1}^n \alpha_{i'}^x \phi\{g(y_i);  x_{i'}^\T \theta^s, 1\}
    }{
    \prod_{i=1}^n  \sum_{i'=1}^n \alpha_{i'}^x \phi\{g(y_i); x_{i'}^\T\mu_\theta, 1 + x_{i'}^\T \Sigma_\theta x_{i'}^\T\}
    }, \quad \{\theta^s\}_{s=1}^S \sim p(\theta)
\end{equation}
where $\phi$ is the Gaussian density function and $\{\alpha_i^x\}_{i=1}^n$ and $g$ are sampled in Algorithm~\ref{alg:joint}. In our simulated data examples, this adjustment has minimal impact on posterior (predictive) inference (see the supplementary material), which suggests that Algorithm~\ref{alg:joint} may be applied directly (i.e., without adjustment) in certain settings.

\subsection{Semiparametric Bayesian quantile regression}\label{sec-sbqr}
We apply model \eqref{trans}--\eqref{mod} and Algorithm~\ref{alg:joint} to improve quantile estimation and posterior inference for Bayesian linear quantile regression. Posterior inference for Bayesian quantile regression is facilitated  by a convenient parameter expansion for an asymmetric Laplace variable: $\epsilon = a_\tau \xi + b_\tau \sqrt{\xi} \eta $, where  $a_\tau = (1-2\tau)/\{\tau(1-\tau)\}$,  $b_\tau = \sqrt{2/\{\tau(1-\tau)\}}$,  $\xi$ and $\eta$ are independent standard exponential and standard Gaussian random variables, respectively. Thus, the regression model \eqref{mod} with $f_\theta(x) = x^\T \theta$ and ALD errors can  be written conditionally (on $\xi$) as a Gaussian linear model. This representation suggests a Gibbs sampling algorithm that alternatively draws $\theta$ from a full conditional Gaussian distribution and each $\xi_i$ independently  from a generalized inverse Gaussian distribution \citep{Kozumi2011}. 

We adapt this  strategy for Algorithm~\ref{alg:joint}. The key step again is to construct $\hat F_{Z\mid X = x_i}$, which is necessary to apply Algorithm~\ref{alg:Fz} and sample $g^* \sim p(g \mid \D_n)$. For computational convenience, we pair an approximation $\hat p(\theta \mid \D_n, g = \hat g) = N(\hat \theta,  \Sigma_{\hat\theta})$ with a parameter expansion of the ALD for $P_{Z \mid \theta, X}$  in \eqref{mod}--\eqref{mod}. Let $\theta \sim N(\mu_\theta, \Sigma_\theta)$ be the prior. The preliminary approximation $\hat p$ may be set at the prior, $\hat \theta = \mu_\theta$ and $\Sigma_{\hat\theta} = \Sigma_{\theta}$, or estimated from the data, e.g., using classical quantile regression for $\{x_i, \hat g(y_i)\}_{i=1}^n$. By marginalizing over this $\hat p$ as in \eqref{Fzx}, the parameter-expanded distribution is $P_{Z \mid X, \xi} = N(x^\T\hat\theta + a_\tau \xi, b_\tau^2 \xi + x^\T \Sigma_{\hat\theta} x)$. Integrating over $\xi \sim \mbox{Exp}(1)$  requires a simple modification of the  estimator from Section~\ref{sec-sblm}: $\hat F_{Z \mid X=x_i}(t) \approx S^{-1} \sum_{s=1}^S \Phi(t; x_i^\T \hat\theta + a_\tau \xi^s, b_\tau^2 \xi^s + x_i^\T \Sigma_{\hat\theta} x_i)$, where $\{\xi^s\}_{s=1}^S \sim \mbox{Exp}(1)$. From our empirical studies, this approximation is accurate even when $S$ is small. The remaining steps to sample $g^* \sim p(g \mid \D_n)$ are straightforward.


Next, we sample $\theta$ using the traditional Gibbs steps, $p(\theta \mid \D_n, g = g^*, \xi  = \xi^*) =  N(Q_\theta^{-1} \ell_\theta, Q_\theta^{-1})$ with $Q_\theta =  X^\T \Sigma_{\xi^*}^{-1} X + \Sigma_\theta^{-1}$,  $\ell_\theta = X^\T \Sigma_{\xi^*}^{-1} \{g^*(y)  - a_\tau \xi^*\} + \Sigma_\theta^{-1}\mu_\theta$, and $\Sigma_{\xi^*} = \mbox{diag}(b_\tau^2 \xi^*)$ with $\xi^* = (\xi_1^*,\ldots,\xi_n^*)^\T$ drawn from the usual independent generalized inverse Gaussian full conditional distributions. As in Section~\ref{sec-sblm}, $\theta$ now includes an intercept; we omit the scale parameter for simplicity, but modifications to include $\sigma$ are available \citep{Kozumi2011}. 

Finally, the predictive sampling step may draw $\tilde z^*(x)$ directly from an ALD or use the  parameter expansion $\tilde z^*(x) \sim N(x^\T \theta^* + a_\tau \tilde \xi, b_\tau^2 \tilde \xi)$ with $\tilde \xi \sim \mbox{Exp}(1)$, and then set $\tilde y^*(x) = (g^*)^{-1}\{\tilde z^*(x)\}$.

Although this version of Algorithm~\ref{alg:joint} is MCMC, we emphasize that the key parameters  $(g, \theta)$ are still blocked efficiently, with $g$ sampled unconditionally on $\theta$.  

\subsection{Scalable semiparametric Gaussian processes}\label{sec-sbgp}
An immensely popular model for \eqref{mod} is the Gaussian process model $f_\theta \sim \mathcal{GP}(m_\theta, \sigma^2K_\theta)$ for mean function $m_\theta$ and covariance function $K_\theta$ parameterized by $\theta \sim p(\theta)$. The nonparametric flexibility of $f_\theta$ is widely useful for spatio-temporal modeling and regression analysis. However, the usual assumption of Gaussian errors is often inappropriate, especially for data that exhibit multimodality, skewness, or kurtosis, or for data with compact or positive support. The transformation \eqref{trans} helps resolve this critical limitation.  However, existing  Bayesian approaches rely on Box-Cox transformations \citep{DeOliveira1997} and often report only posterior modes \citep{rios2018learning,Lin2020} or variational approximations   \citep{lazaro2012bayesian}. 

Algorithm~\ref{alg:joint} offers a solution. Once again, the critical step is to construct $\hat F_{Z \mid X=x_i}$  to apply Algorithm~\ref{alg:Fz} and  sample $g^*\sim p(g \mid \D_n)$. To facilitate direct and feasible computation, we prioritize the uncertainty from $(g, f_\theta, \tilde y(x))$ and fix  $\theta$ at an optimal value; generalizations are discussed below. For inputs $\{x_i, \hat g(y_i)\}_{i=1}^n$, we  compute the maximum likelihood estimator (MLE) $\hat \theta$ for $\theta$ and the (conditional) posterior distribution for the regression function, 
$ \hat p(f_{\hat\theta} \mid \D_n, g = \hat g)  = N(\hat f_{\hat\theta}, \sigma^2  \Sigma_{f_{\hat\theta}})$, where $\hat f_{\hat\theta} = \{\hat f_{\hat\theta}(x_i)\}_{i=1}^n$ are the point predictions  at $\{x_i\}_{i=1}^n$ given data $\{x_i, \hat g(y_i)\}_{i=1}^n$ and  $\Sigma_{f_{\hat\theta}} = (K_{\hat\theta}^{-1} + I_n)^{-1}$ with $K_{\hat\theta} = \{K_{\hat\theta}(x_i, x_{i'})\}_{i,i'=1}^n$. Importantly, these are standard quantities in Gaussian process estimation, and thus we can leverage state-of-the-art algorithms and software. Finally, the critical term for sampling $g^* \sim p(g \mid \D_n)$ is $\hat F_{Z \mid X = x_i}(t) = \Phi[t; \hat f_{\hat \theta}(x_i),  \sigma^2\{1 + (\Sigma_{f_{\hat\theta}})_{ii}\}]$, where $\sigma = 1$ is fixed for identifiability.

The remainder of Algorithm~\ref{alg:joint} is straightforward. Given $g^* \sim p(g \mid \D_n)$ and reintroducing the scale $\sigma$ for robustness, we sample $f_{\hat \theta}^* \sim \hat p(f_{\hat\theta} \mid \{x_i, g^*(y_i)\}_{i=1}^n)  = N(\hat f_{\hat\theta}, \sigma^2\Sigma_{f_{\hat\theta}})$, where $\hat f_{\hat\theta}$ are now the point predictions at $\{x_i\}_{i=1}^n$ given data $\{x_i, g^*(y_i)\}_{i=1}^n$. Sampling $\sigma$ may proceed using similar strategies as in  Section~\ref{sec-sblm}.  The predictive sampling step is $\tilde z^*(x) \sim N(f_{\hat \theta}^*(x_i), \sigma^2)$  and  $\tilde y^*(x) = (g^*)^{-1}\{\tilde z^*(x)\}$; modifications for out-of-sample predictive draws are readily available.

If instead we wish  to also account for the uncertainty of $\theta$, there are two main modifications required. First, given an approximate posterior  $\hat p(\theta \mid \D_n)$, we modify the key term in Algorithm~\ref{alg:Fz}: $\hat F_{Z \mid X = x_i}(t) \approx S^{-1}\sum_{s=1}^S \Phi[t; \hat f_{\theta^s}(x_i),  \sigma^2\{1 + (\Sigma_{f_{\theta^s}})_{ii}\}]$, where  $\{\theta^s\}_{s=1}^S \sim \hat p(\theta \mid \D_n)$. Second, we must sample $\theta^* \sim p(\theta \mid \D_n, g = g^*)$ and replace $\hat \theta$ with $\theta^*$ in the sampling steps for $f_\theta$. Of course, these approximate and conditional posterior distributions for $\theta$ will be specific to the mean function $m_\theta$ and covariance function $K_\theta$ in the Gaussian process model. 

Yet even when $\theta = \hat \theta$ is fixed, posterior sampling of $f_\theta$ is a significant computational burden. We use a fast approximation that  bypasses these sampling steps. Specifically, we  fix $\theta= \hat\theta$, $f_{\hat\theta} = \hat f_{\hat \theta}$, and $\sigma = \hat \sigma$ at their MLEs from the initialization step using data $\{x_i, \hat g(y_i)\}$, where $\hat \sigma$ is included for robustness akin to Section~\ref{sec-approx}. The key term in Algorithm~\ref{alg:Fz} is now $\hat F_{Z \mid X = x_i}(t) = \Phi[t; \hat f_{\hat \theta}(x_i),  \hat \sigma^2\{1 + (\Sigma_{f_{\hat\theta}})_{ii}\}]$. Bypassing the sampling steps for $(\theta, f_\theta, \sigma)$,  the predictive sampling step  is now simply $\tilde z^*(x) \sim N(\hat f_{\hat \theta}(x), \hat \sigma^2)$  and  $\tilde y^*(x) = (g^*)^{-1}\{\tilde z^*(x)\}$.

Relative to point estimation for (untransformed) Gaussian processes, this latter approach requires only one additional optimization step and a series of simple and fast sampling steps. The fully Bayesian model for the transformation $g$ is especially important here: it helps correct not only for model inadequacies that may arise from a  Gaussian model for $y$---for example, if the errors are multimodal or skewed or if $\mathcal{Y}$ is compact---but also for the approximations obtained by fixing parameters at point estimates.  This strategy is evaluated empirically in Section~\ref{sec-lidar}.

\section{Empirical results}\label{sec-emp}

\subsection{Simulation study for semiparametric Bayesian linear regression}\label{sec-lm}
We evaluate the proposed semiparametric Bayesian linear models for prediction and inference using simulated data. Data are generated from a transformed linear model  with $n$ observations and $p$ covariates, where the covariates are marginal standard Gaussian with $\mbox{corr}(x_{ij}, x_{ij'}) = (0.75)^{\vert j - j'\vert}$ and randomly permuted columns. Latent data are simulated from a Gaussian linear model with $p/2$ true regression coefficients set to one and the rest set to zero, and unit error standard deviation (0.25 and 1.25 produced similar results). 

We consider three inverse transformation functions, which are applied to these latent data (after centering and scaling) to generate $y$. These transformations determine both the support $\mathcal{Y}$ and the complexity of the link between the linear term  and $y$.  First, we induce an approximate Beta marginal distribution with $g^{-1}(t) = F_{{\rm Beta}(0.1, 0.5)}^{-1}\{\Phi(t)\}$, which yields  $\mathcal{Y}=[0,1]$ with many $y$ values near zero   (\texttt{beta}). Second, we generate a monotone and locally linear function by simulating 10 increments  identically from  a standard exponential distribution at equally-spaced points on $[-3,3]$ and linearly interpolating the cumulative sums, which produces positive  data   $\mathcal{Y} = \mathbb{R}^+$ with a nontrivial transformation (\texttt{step}). Third, we specify an inverse (signed) Box-Cox function with $\lambda = 0.5$ (see below), which corresponds to a (signed) square-root transformation  and thus $\mathcal{Y} = \mathbb{R}$ (\texttt{box-cox}).   For each simulation, a testing dataset $(X^{test}, y^{test})$ with 1000 observations is generated independently and identically. This process is repeated 100 times for each inverse transformation function and  $(n,p) \in \{(50, 10), (200,50)\}$. 

We evaluate several Bayesian approaches. In each case, the linear coefficients are assigned a $g$-prior with $\mu_\theta = 0$ and $\Sigma_\theta = n \sigma^2 (X^\T X)^{-1}$. For the proposed approach, we implement  Algorithm~\ref{alg:joint} for the semiparametric Bayesian linear model as in Section~\ref{sec-sblm} using the Laplace approximation for $\hat p(\theta \mid \D_n)$ (\texttt{sblm}) or the prior  (\texttt{sblm(prior)}). We also include a simplification that fixes the transformation at the initialization $\hat g$, which does not account for the uncertainty in $g$ (\texttt{sblm(fixed)}). 
For benchmarking, we include a Bayesian linear model without a transformation (\texttt{blm}),  with a Box-Cox transformation $g(t; \lambda) = \{\mbox{sign}(t)\vert t\vert^\lambda - 1\}/\lambda$ (\texttt{blm(box-cox)}), and with a nonparametric spline model for the transformation (\texttt{blm(spline)}). 
The Box-Cox model uses the prior $\lambda \sim N(0.5, 0.5^2)$ truncated to $(0,2)$, which centers the unknown transformation at the (signed) square-root, and samples the joint posterior using a slice-within-Gibbs sampler that alternates between a slice sampler for $[\lambda \mid \D_n, \theta, \sigma]$ \citep{neal2003slice} and the sampling steps for $[\theta, \sigma \mid \D_n, g]$ from Section~\ref{sec-sblm}. \texttt{blm(spline)} specifies an I-spline basis expansion with monotonicity constraints for $g$, and uses a Metropolis-within-Gibbs sampler with a robust adaptive Metropolis (RAM) sampler \citep{vihola2012robust}  for $[g \mid \D_n, \theta, \sigma]$; 
details are in the supplementary material. 
For each implementation, we generate and store 1000 samples from the posterior of $\theta$ and the joint posterior predictive distribution on the testing data $\tilde y(X^{test})$. 

For a non-Bayesian competitor, we include a conditional transformation model (\texttt{ctm}; \citealp{hothorn2014conditional}) implemented in the \texttt{R} package \texttt{mlt} \citep{hothorn2020most}. We fit the \texttt{ctm}   $F_{Y \mid \theta, X=x}(t) = F_0\{h(t \mid \theta, x)\}$ with reference $F_0 = \Phi$ and conditional transformation $h(t \mid \theta, x) =  a_{BS}^\T(t)\theta_0 - x^\T\theta$, where $a_{BS}$ contains 9 Bernstein polynomial basis functions.  
From the fitted \texttt{ctm}, we compute the 5th and 95th quantiles at $X^{test}$ to generate 90\% prediction intervals for the testing data.

First, we evaluate predictive performance by comparing the width and empirical coverage of the 90\% out-of-sample prediction intervals (Figure~\ref{fig:sims-lm}); similar trends are observed for continuous ranked probability scores (see the supplement).  Most notably, both \texttt{sblm} and \texttt{sblm(prior)} are precise and well-calibrated: the prediction intervals are narrow and achieve approximately the nominal coverage. As expected,  \texttt{sblm(fixed)} produces narrower intervals but below-nominal coverage, which shows the importance of accounting for the uncertainty of $g$. 
The competing methods \texttt{blm}, \texttt{blm(box-cox)}, and \texttt{blm(spline)}  fail to provide both precision and calibration, even for the true \texttt{box-cox} design. Finally, \texttt{ctm} (not shown) achieves the nominal coverage in each case, but produces intervals that are substantially wider than all other methods.

\begin{figure}[h]
\centering
\includegraphics[width=.32\textwidth]{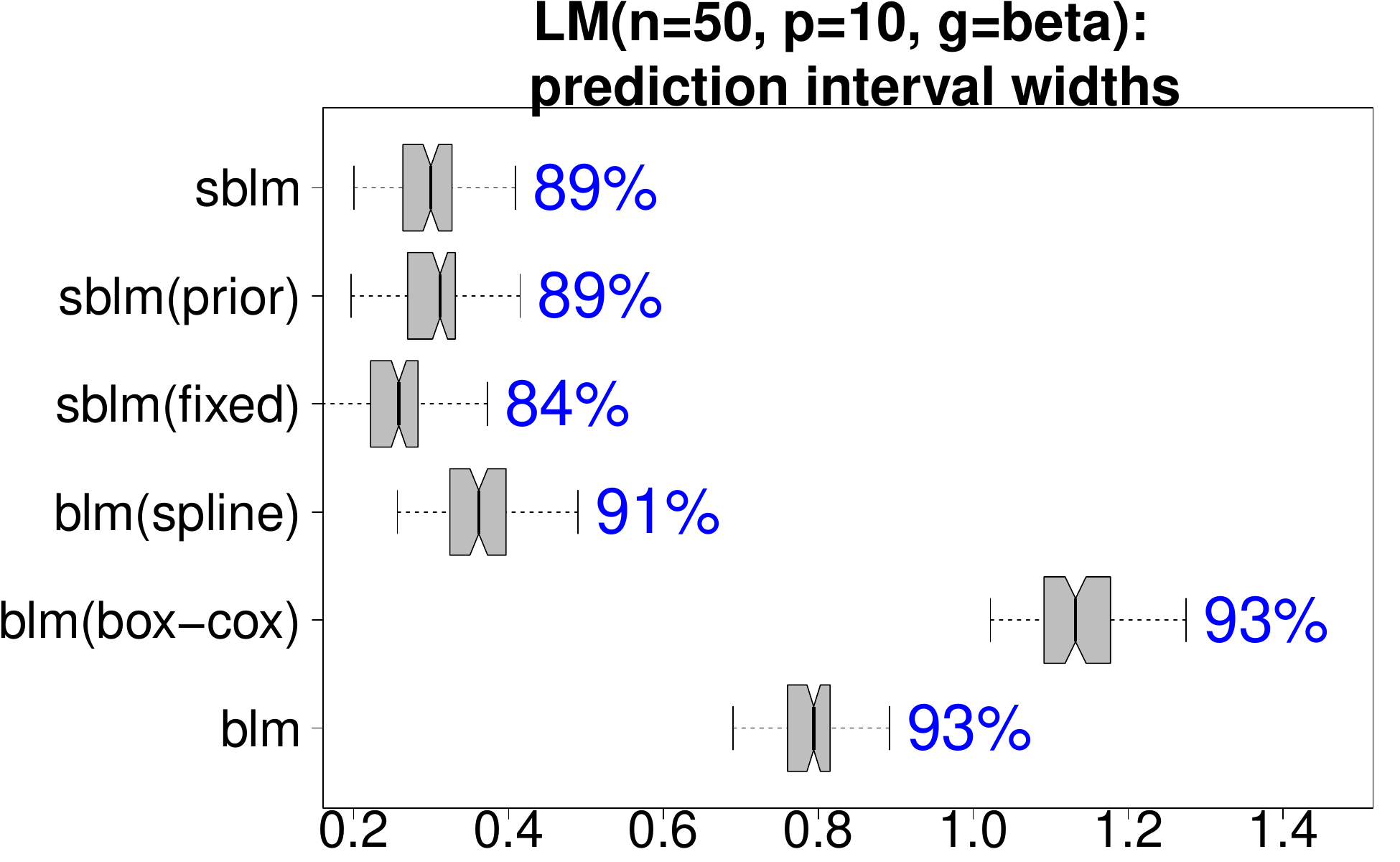}
\includegraphics[width=.32\textwidth]{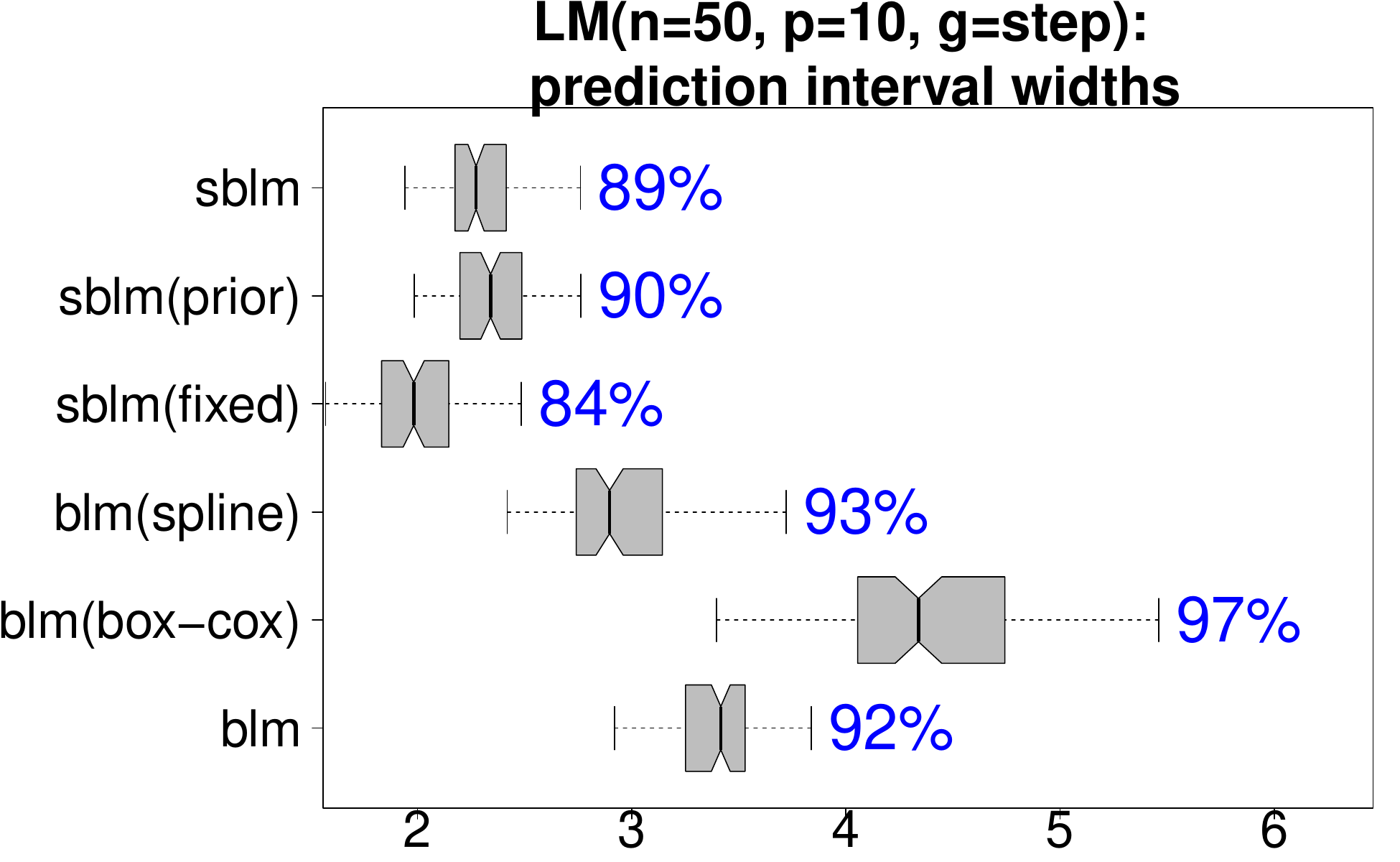}
\includegraphics[width=.32\textwidth]{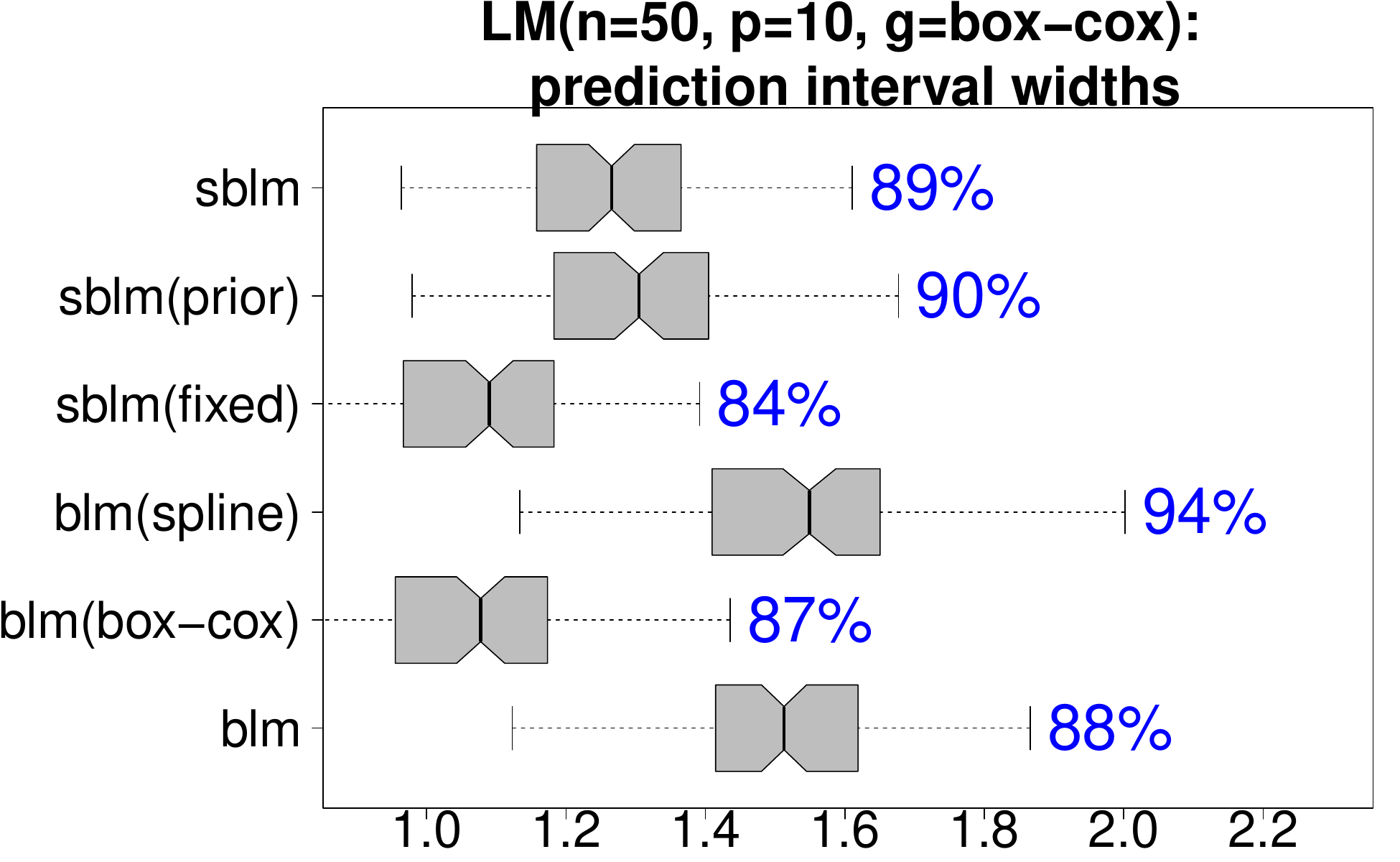}
\includegraphics[width=.32\textwidth]{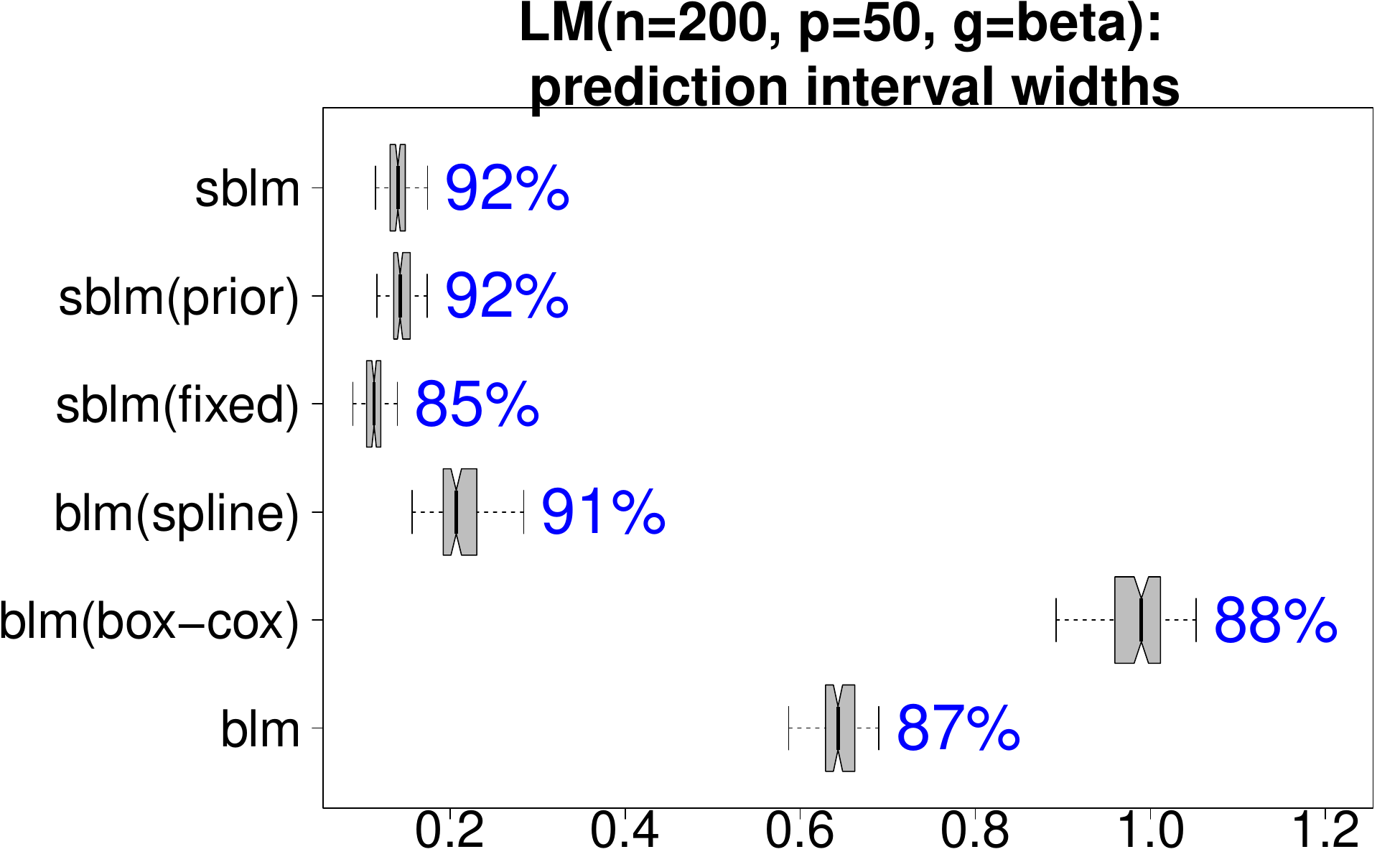}
\includegraphics[width=.32\textwidth]{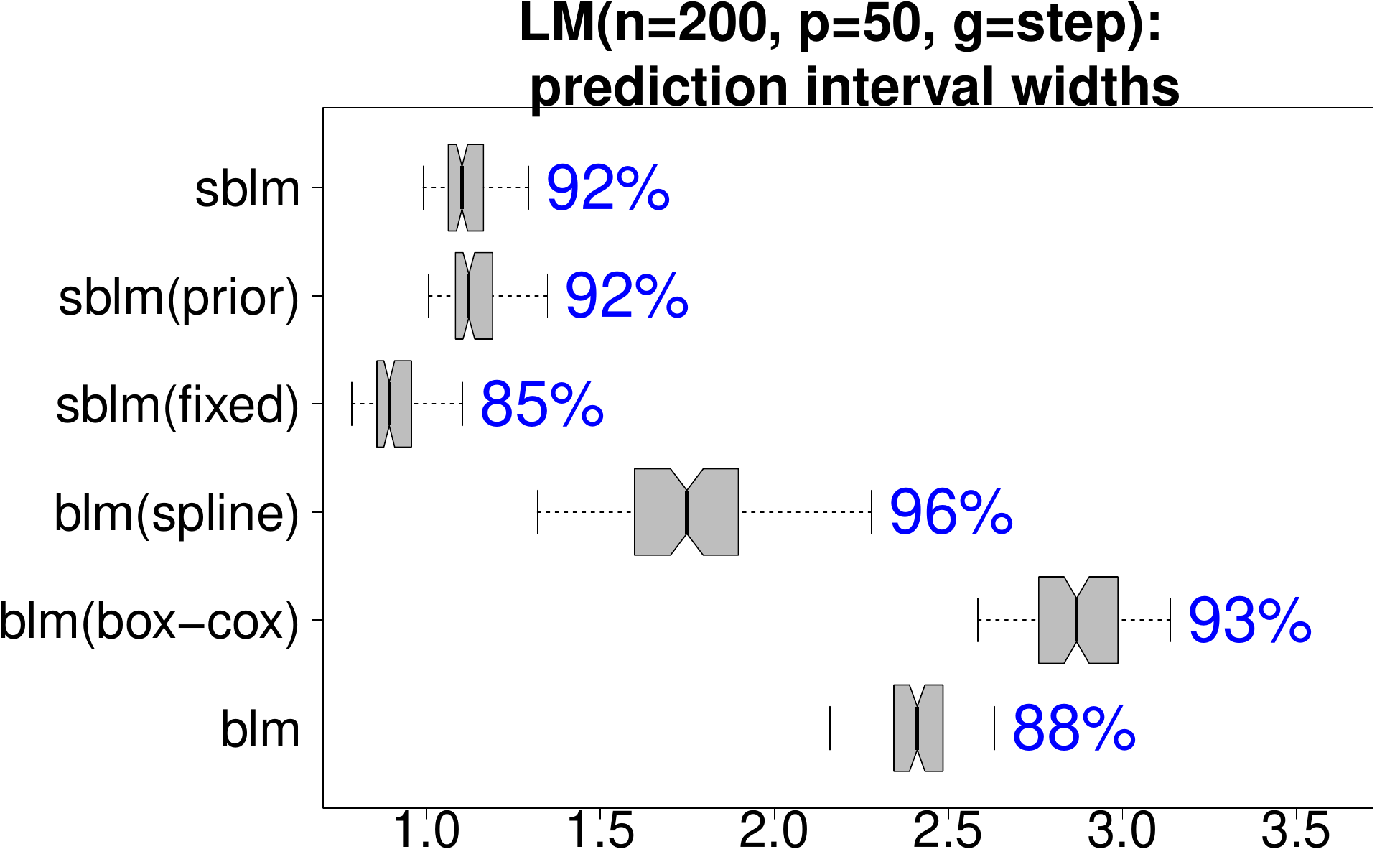}
\includegraphics[width=.32\textwidth]{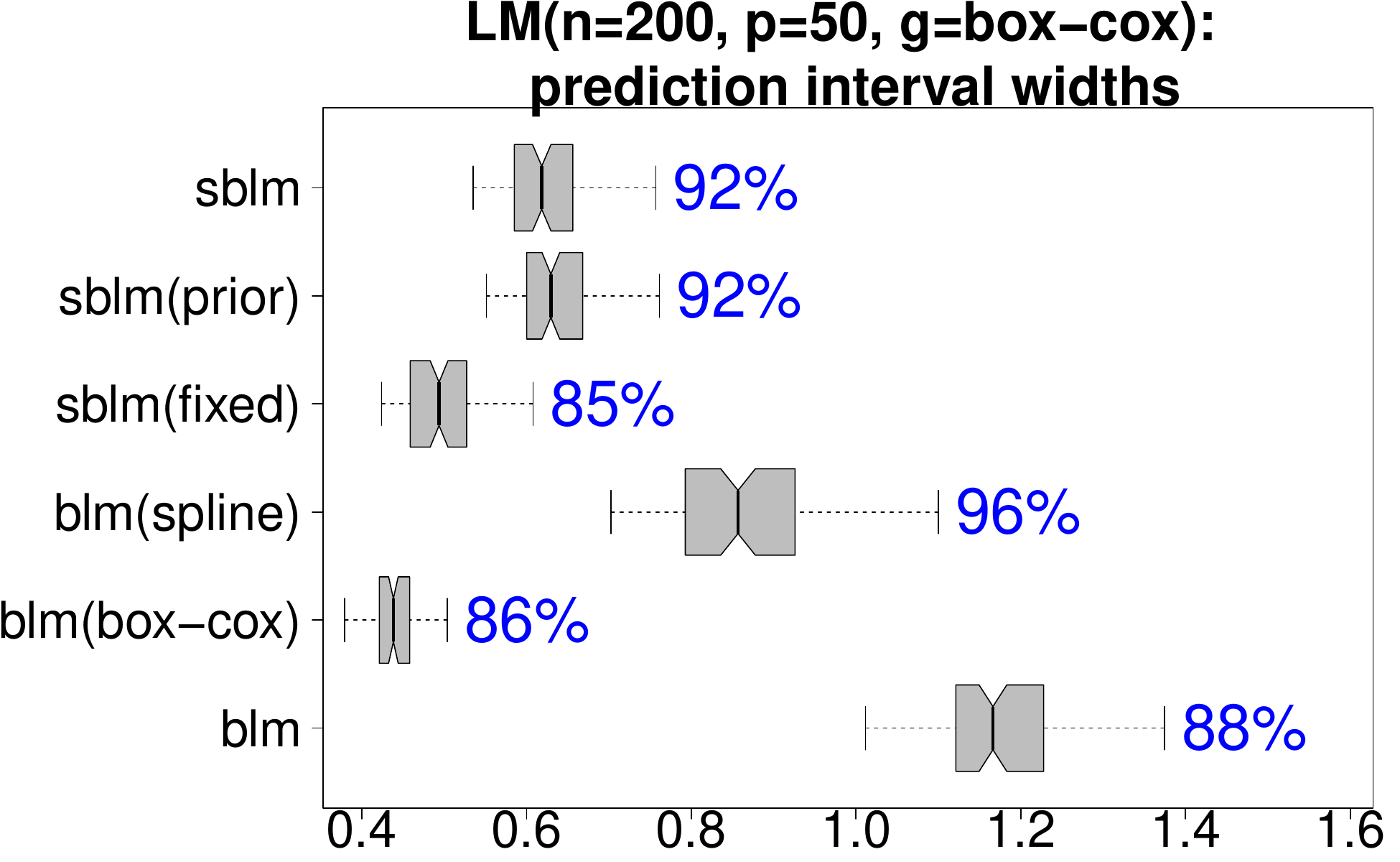}
 \caption{90\% prediction interval widths and  empirical coverage (annotated) for out-of-sample prediction with semiparametric Bayesian linear models across various designs. 
 The proposed semiparametric models provide narrow intervals that achieve approximately the correct nominal coverage, with dramatic gains over competing methods for more complex transformations.  \texttt{ctm} average interval widths were much larger ($\approx 4$ for all designs) and are omitted.
\label{fig:sims-lm}}
\end{figure}

Next, we evaluate inference for the regression coefficients $\theta$ using variable selection. Although the scale of $\theta$ depends on the transformation---and  thus differs among competing methods---the determination of whether each $\theta_j \ne 0$ is more comparable. Here, we select variables if the 95\% highest posterior density interval for $\theta_j$ excludes zero. The true positive and negative rates are averaged  across 100 simulations and presented in Table~\ref{tab:select}. 
The transformation is critical: for the \texttt{beta} and \texttt{step} designs, the proposed \texttt{sblm} methods offer a massive increase in the power to detect true effects without incurring more false discoveries. Remarkably, both \texttt{sblm} and \texttt{sblm(prior)} are highly robust to the true transformation and improve rapidly with the sample size, even as $p$ grows. By comparison,  \texttt{blm}, \texttt{blm(box-cox)}, and  \texttt{blm(spline)} are excessively conservative in their interval estimates for the regression coefficients  and lack the same improvements with larger $n$. Additional evaluations for point estimation of $\theta$ are in the supplementary material.

\begin{table}[h] 
\scriptsize	
\centering
\begin{tabular}{c | c | c | c | c | c | c | c }
 $(n, p)$ & design & \texttt{blm} & \texttt{blm(box-cox)} & 
 \texttt{blm(spline)} &
 \texttt{sblm(fixed)} & \texttt{sblm(prior)} &\texttt{sblm} \\ 
 \hline 
\multirow{3}{*}{$(50, 10)$}  
    & \texttt{beta} & $0.18 \quad 0.96$  & $0.13\quad 0.96$  & $0.62 \quad 0.98$ & $0.85\quad 0.94$ & $0.75\quad 0.98 $     & $0.76\quad 0.99$    \\
    & \texttt{step}   &    $0.51 \quad 0.98$ & $0.38 \quad 0.99$ & $0.55 \quad 0.99$ & $0.85 \quad 0.94$ & $0.75 \quad 0.98$ & $0.75 \quad 0.99$   \\
    & \texttt{box-cox}  &  $0.66 \quad 0.94 $ & $ 0.89 \quad 0.96$ & $0.61 \quad 0.99$ & $0.85 \quad 0.93$ & $0.75 \quad 0.98$ & $0.76 \quad 0.99$   \\  
                           \hline 
\multirow{3}{*}{$(200, 50)$}  
    & \texttt{beta}   & $0.18 \quad 0.92$  & $0.14\quad 0.92$  & $0.82 \quad 0.96$ & $1.00\quad 0.96$ & $0.99 \quad 0.99$ & $0.99 \quad 0.99$   \\
    & \texttt{step}   &    $0.52 \quad 0.93$ & $0.46 \quad 0.96$ & $0.81 \quad 0.99$ & $1.00 \quad 0.95$ & $0.99 \quad 0.99$ & $0.99 \quad 0.99$   \\
    & \texttt{box-cox}   &    $0.58 \quad 0.91 $ & $1.00 \quad 0.97 $ & $0.88 \quad 0.99$ & $1.00 \quad 0.95$ & $0.99 \quad 0.99$ & $0.99 \quad 0.99$   \\
\end{tabular}
\caption{True positive rates and true negative rates for variable selection across simulation designs for the transformed linear model. When the transformation is more complex, the proposed semiparametric Bayesian linear models deliver a massive increase in power to detect true effects. 
\label{tab:select}}
\end{table}

\begin{figure}[h]
\centering
\includegraphics[width=.32\textwidth]{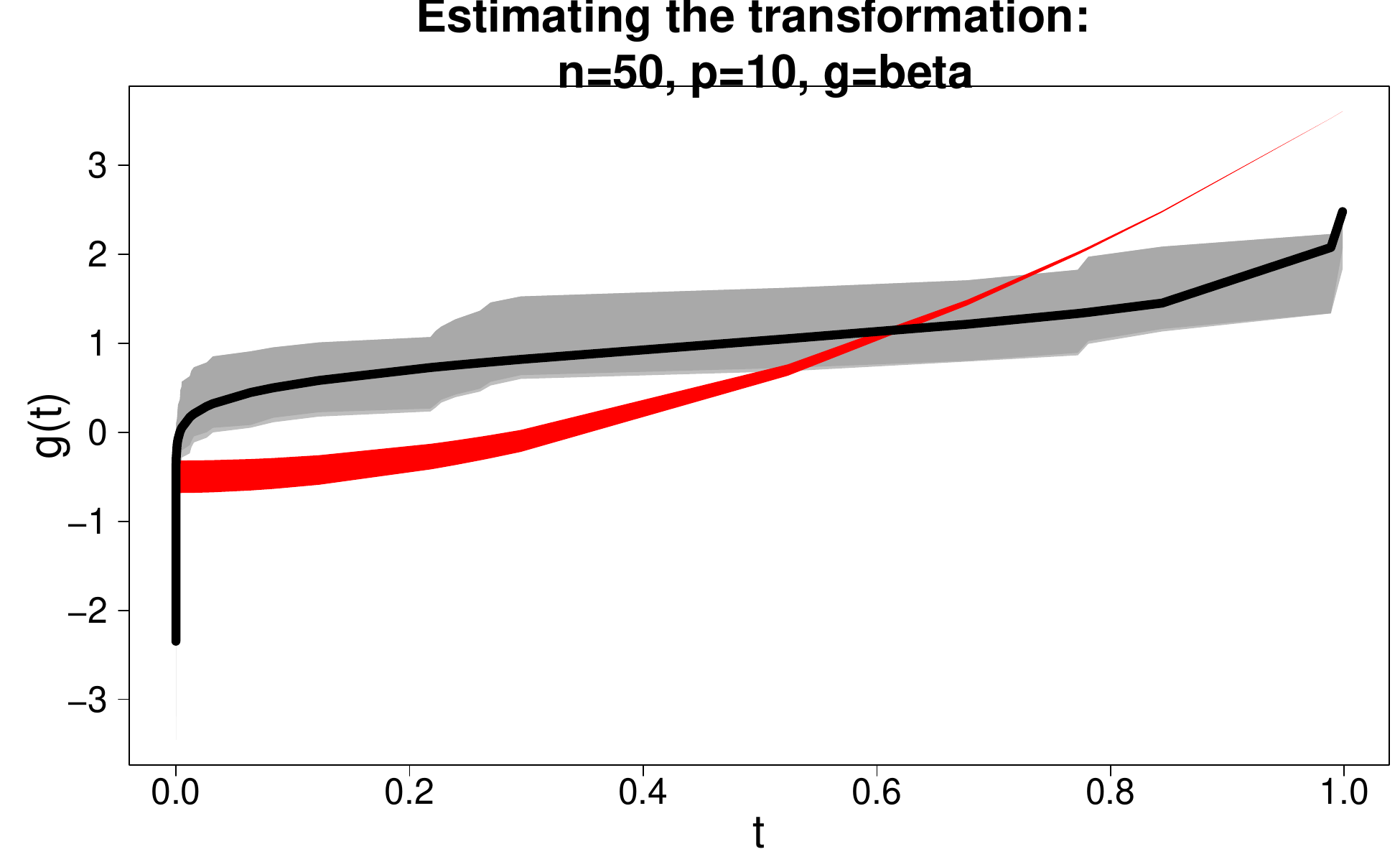}
\includegraphics[width=.32\textwidth]{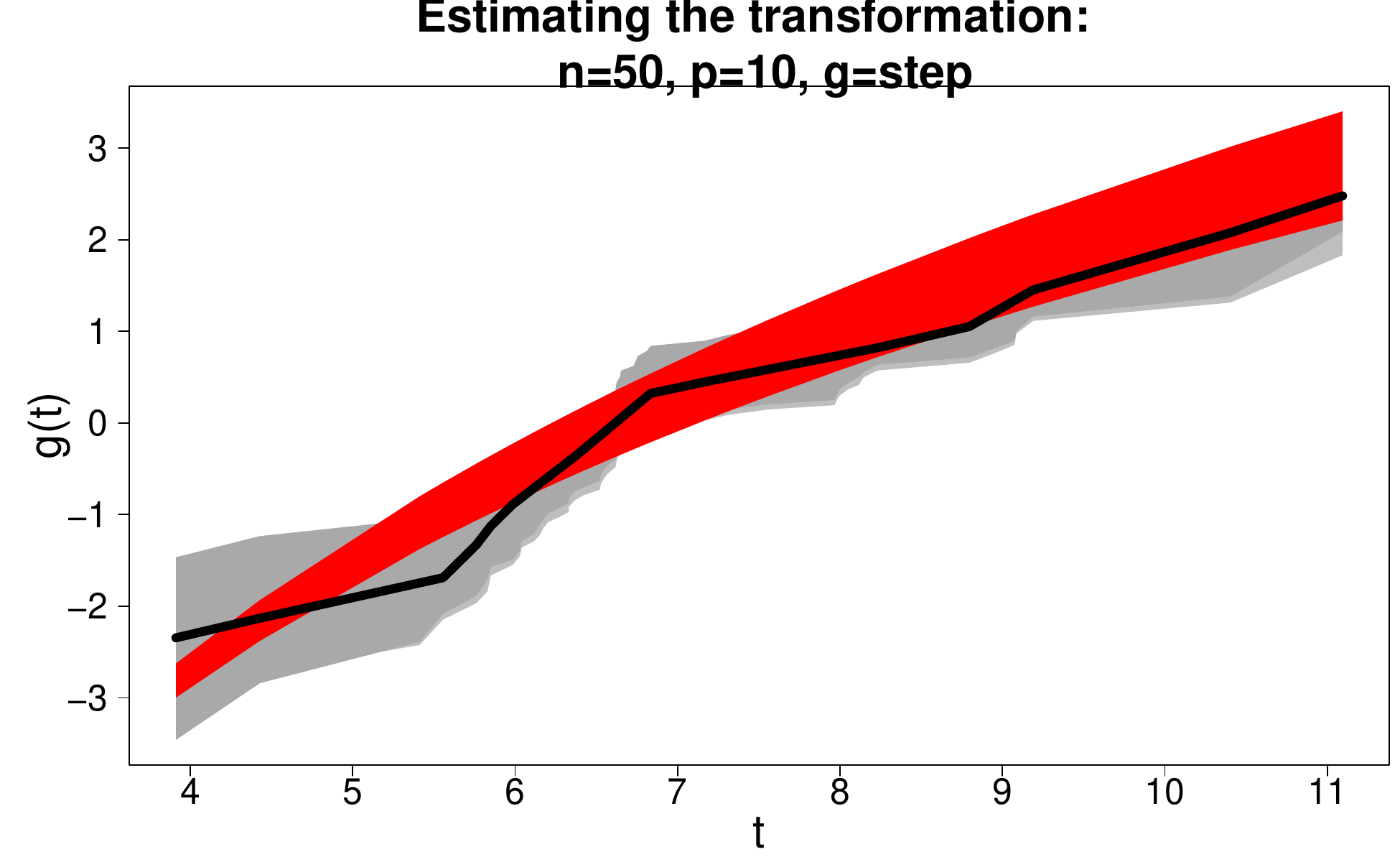}
\includegraphics[width=.32\textwidth]{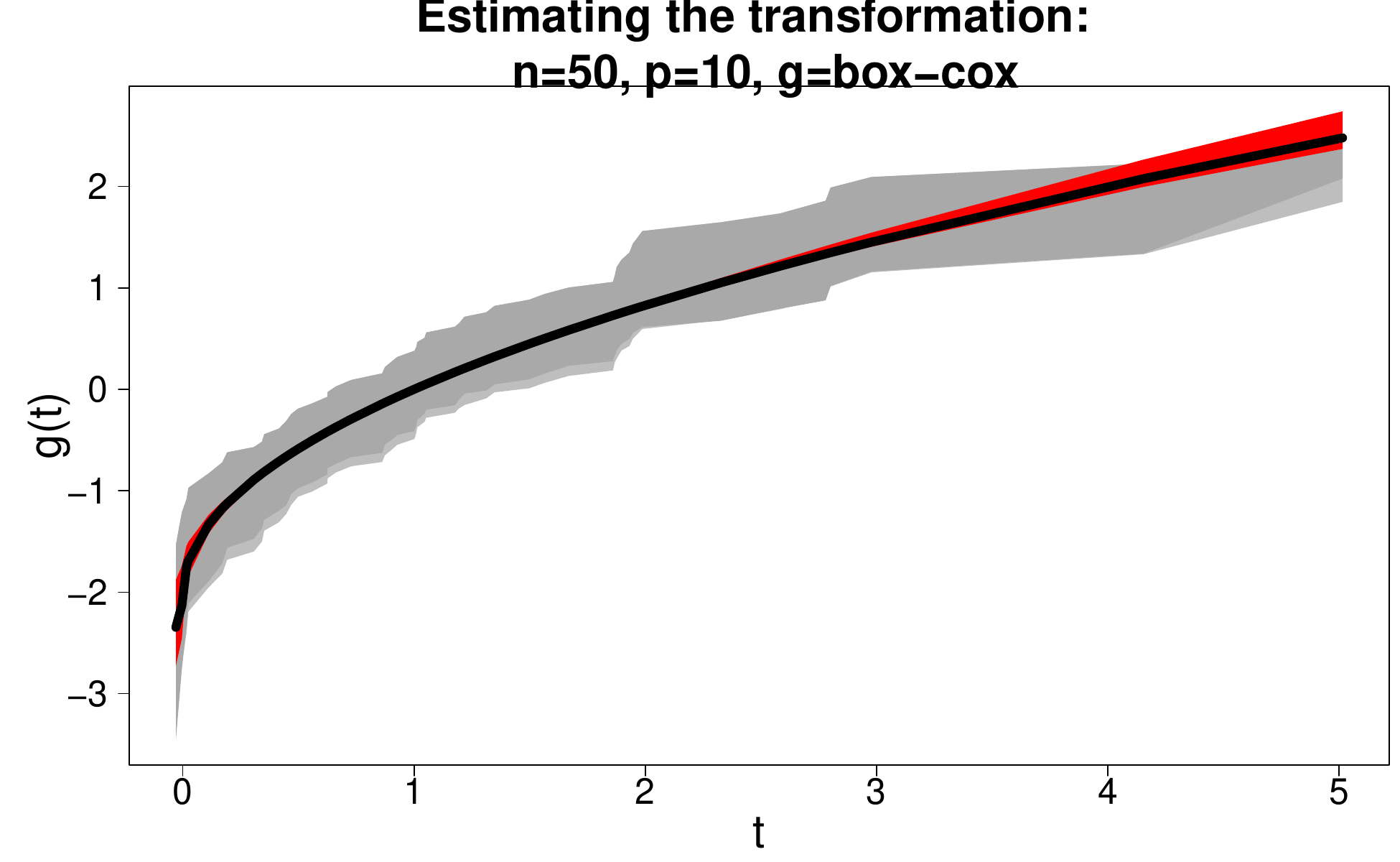}
\includegraphics[width=.32\textwidth]{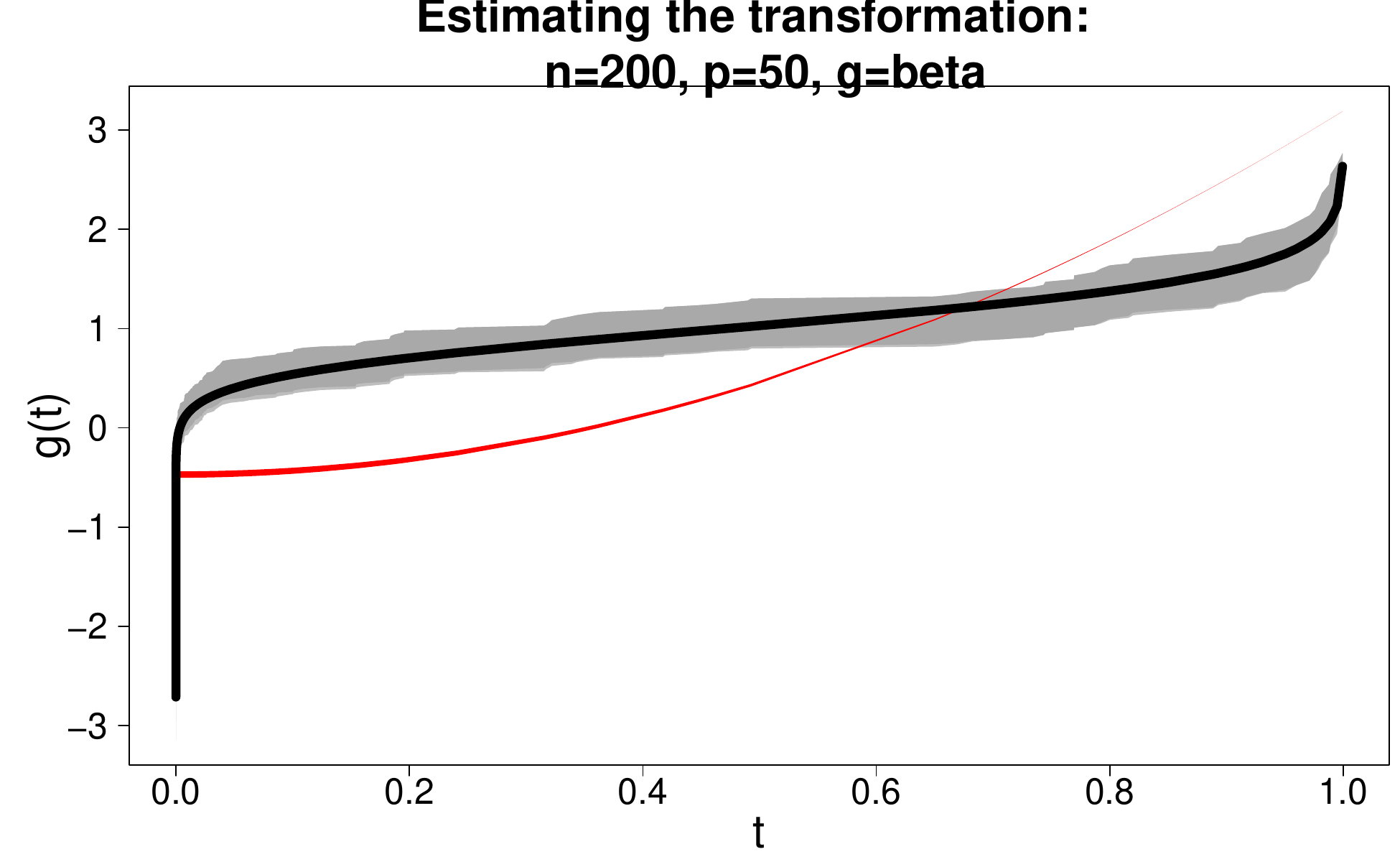}
\includegraphics[width=.32\textwidth]{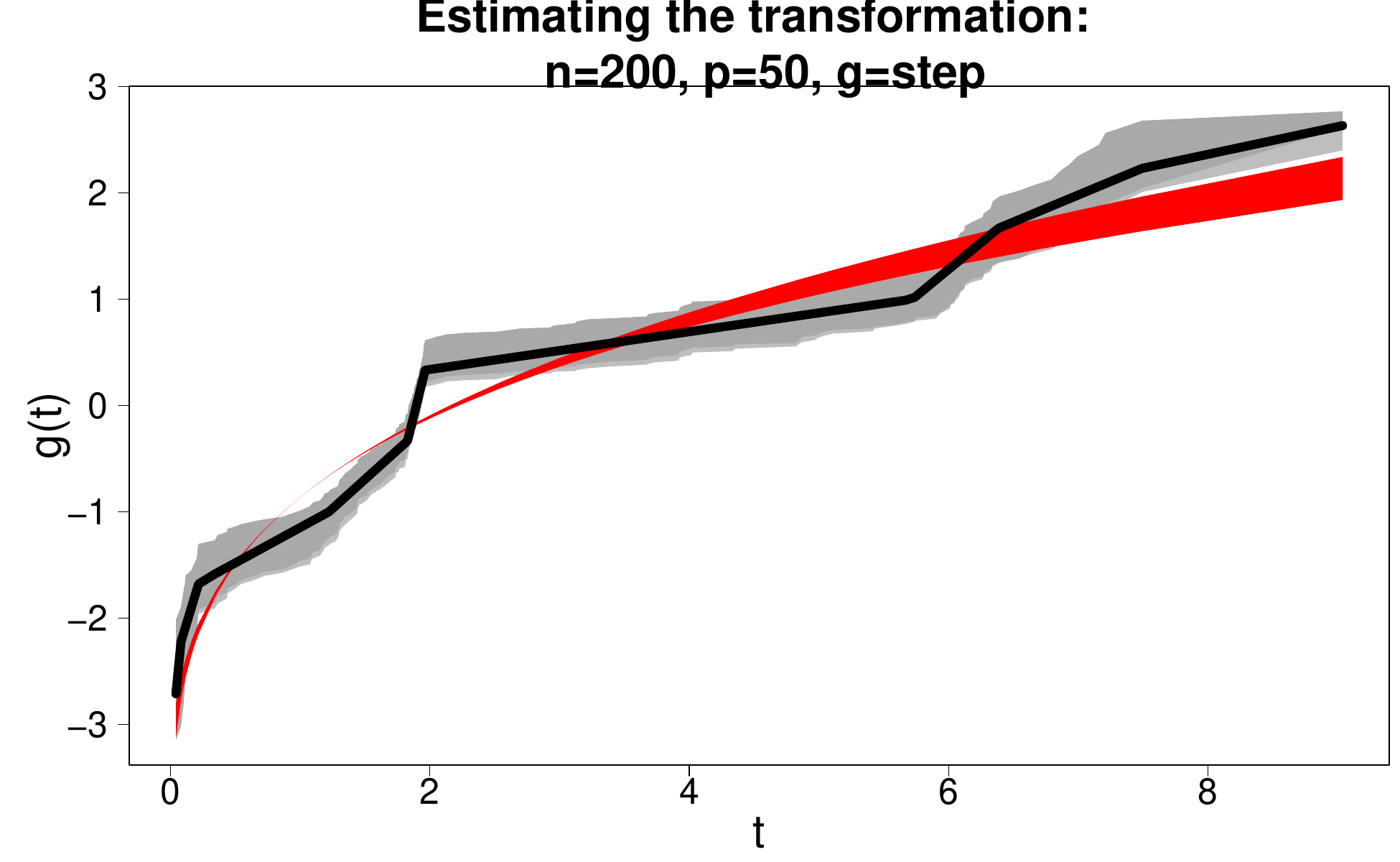}
\includegraphics[width=.32\textwidth]{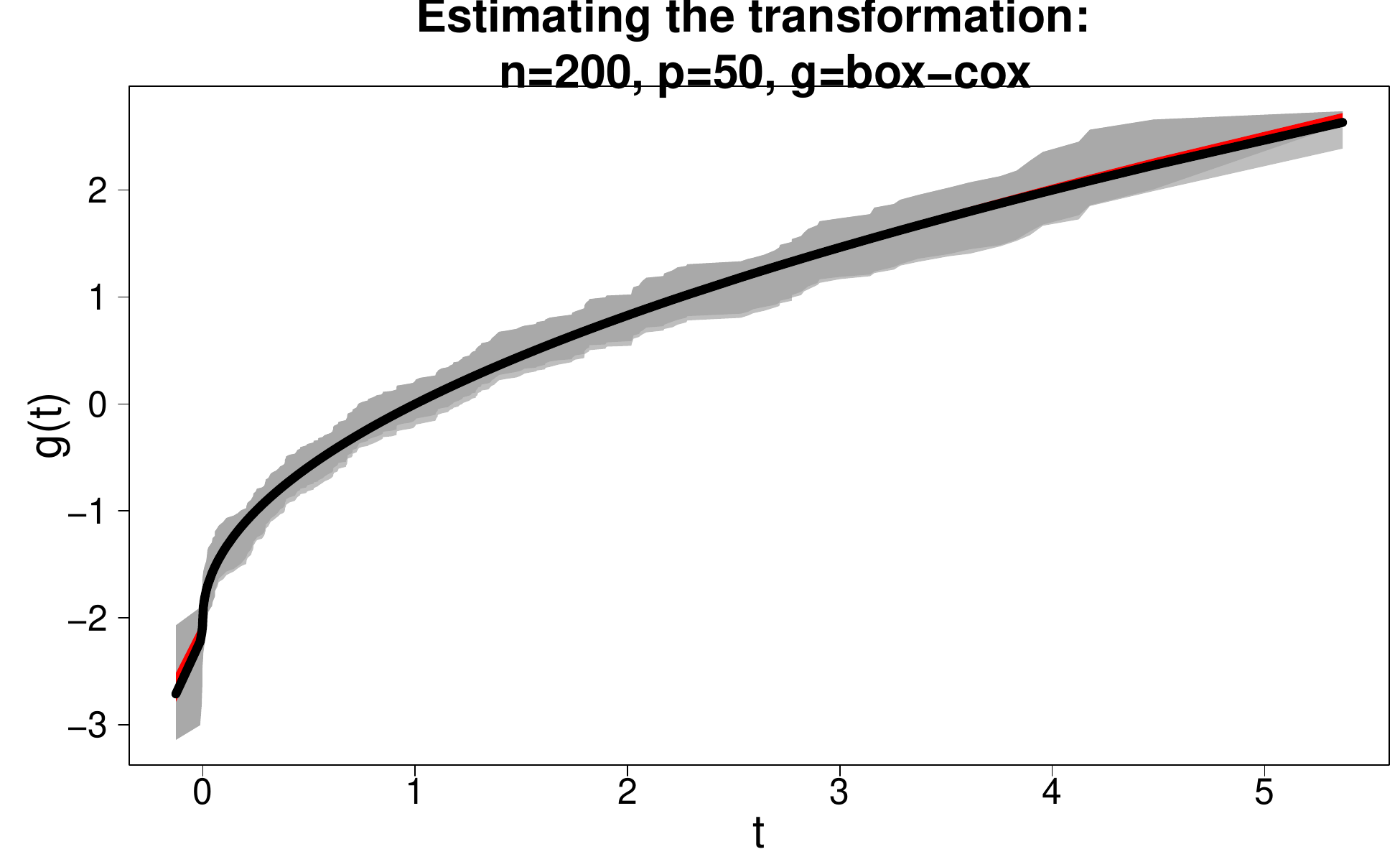}
 \caption{95\% credible intervals for the transformation $g$ using \texttt{sblm} (dark gray), \texttt{sblm(prior)} (light gray), and \texttt{blm(box-cox)} (red) compared to the truth (black). Both of the proposed semiparametric models capture all variety of shapes and concentrate around the truth as $n$ grows.
\label{fig:sims-trans}}
\end{figure}

We highlight the ability of the \texttt{sblm} and \texttt{sblm(prior)} to infer the various transformations on $\Y = [0,1]$, $\Y = \mathbb{R}^+$, and $\Y = \mathbb{R}$. Figure~\ref{fig:sims-trans} presents 95\% pointwise credible intervals for $g$ under these models and \texttt{blm(box-cox)} for a single simulated dataset from each design. The transformations are rescaled such that the posterior means and the true transformations are centered at zero with unit scale, and thus are comparable. Most notably, \texttt{sblm} and \texttt{sblm(prior)}  are virtually indistinguishable 
and successfully concentrate around each true transformation as $n$ grows. By comparison, \texttt{blm(box-cox)} is insufficiently flexible and substantially underestimates the uncertainty about $g$. Additional results for  $n\in\{500, 2000\}$ are in the supplementary material.

Remarkably, using the prior distribution  for $\hat p(\theta \mid \D_n)$ in \eqref{Fzx} and Algorithm~\ref{alg:Fz} (\texttt{sblm(prior)}) performs nearly identically to the data-driven Laplace approximation (\texttt{sblm}) for prediction of $y^{test}$, selection of coefficients $\theta_j \ne 0$, and estimation of the transformation $g$. This suggests that our approach, including the  approximation strategy  from Section~\ref{sec-approx}, is robust to the choice of $\hat p(\theta \mid \D_n)$. Clearly, the ability to substitute $p(\theta)$ is highly beneficial, as it is always available and requires no additional computations or tuning. 

We briefly mention computing performance. Applying Algorithm~\ref{alg:joint} as described in Section~\ref{sec-sblm}, the joint Monte Carlo sampler for    $(\theta, g, \tilde y(X^{test}))$ requires about 3.5 seconds per 1000 samples for the larger $n=200$, $p=50$ design (using \texttt{R} on a MacBook Pro, 2.8 GHz Intel Core i7).  
Because these are Monte Carlo (not MCMC) algorithms, no convergence diagnostics, burn-in periods, or  inefficiency factors are needed. By comparison, the MCMC-based \texttt{blm(spline)} requires about 5 seconds for 1000 samples, including a burn-in/adaptation period of 10000 draws. Yet even though RAM is tuned adaptively,   the minimum effective sample sizes  are only about 150 for $\theta$ (excluding the intercept) and well below 50 for $g(y)$. Further computational comparisons with the MCMC-based alternatives are in the supplementary material.

\subsection{Simulation study for semiparametric Bayesian quantile regression}\label{sims-qr}
We modify the simulation design from Section~\ref{sec-lm} to evaluate the proposed semiparametric approach for quantile regression. First, the latent data are generated from  $z_{i} = x_{i}^\T \theta_{true}(1  + \epsilon_{i})$ with $\epsilon_{i} \sim N(0,1)$, which introduces heteroskedasticity. Heteroskedasticity is a common motivation for quantile regression,  since it often leads  to different conclusions compared to mean regression. Second, the inverse transformation is simply the  identity. Thus, the data-generating process does not implicitly favor the transformed regression model \eqref{trans}--\eqref{mod}. 

We implement the proposed semiparametric Bayesian quantile regression  from Section~\ref{sec-sbqr} with similar variations for inferring $g$ as in Section~\ref{sec-lm}. To specify $\hat p(\theta \mid \D_n)$ in \eqref{Fzx} and Algorithm~\ref{alg:Fz}, we consider both a Laplace approximation  using classical quantile regression with bootstrap-based covariance estimate from the  \texttt{R} package \texttt{quantreg} (\texttt{sbqr}) 
and the prior $p(\theta)$ (\texttt{sbqr(prior)}). We also consider the simplification with the transformation fixed at $\hat g$ in \eqref{trans-cdf} (\texttt{sbqr(fixed)}). For  comparisons, we include  Bayesian quantile regression (\texttt{bqr}) without the transformation, which otherwise uses the same sampling steps as in Section~\ref{sec-sbqr}, and frequentist quantile regression  (\texttt{qr}) using default settings in \texttt{quantreg}. The Bayesian models use the same $g$-prior with $\mu_\theta = 0$ and $\Sigma_\theta = n \sigma^2 (X^\T X)^{-1}$. The models are estimated for quantiles $\tau \in\{0.05,  0.25, 0.50\}$; performance is comparable for large quantiles ($1-\tau$) and the  results for $\tau =0.05$ and $\tau = 0.10$ are similar. Quantile estimates on the testing  data  $X^{test}$ are computed using $X^{test}\hat \theta$ for \texttt{qr} and \texttt{bqr} and the posterior mean of 
$g^{-1}(X^{test}\theta)$ for the semiparametric  methods.

\begin{figure}[h]
\centering
\includegraphics[width=.32\textwidth]{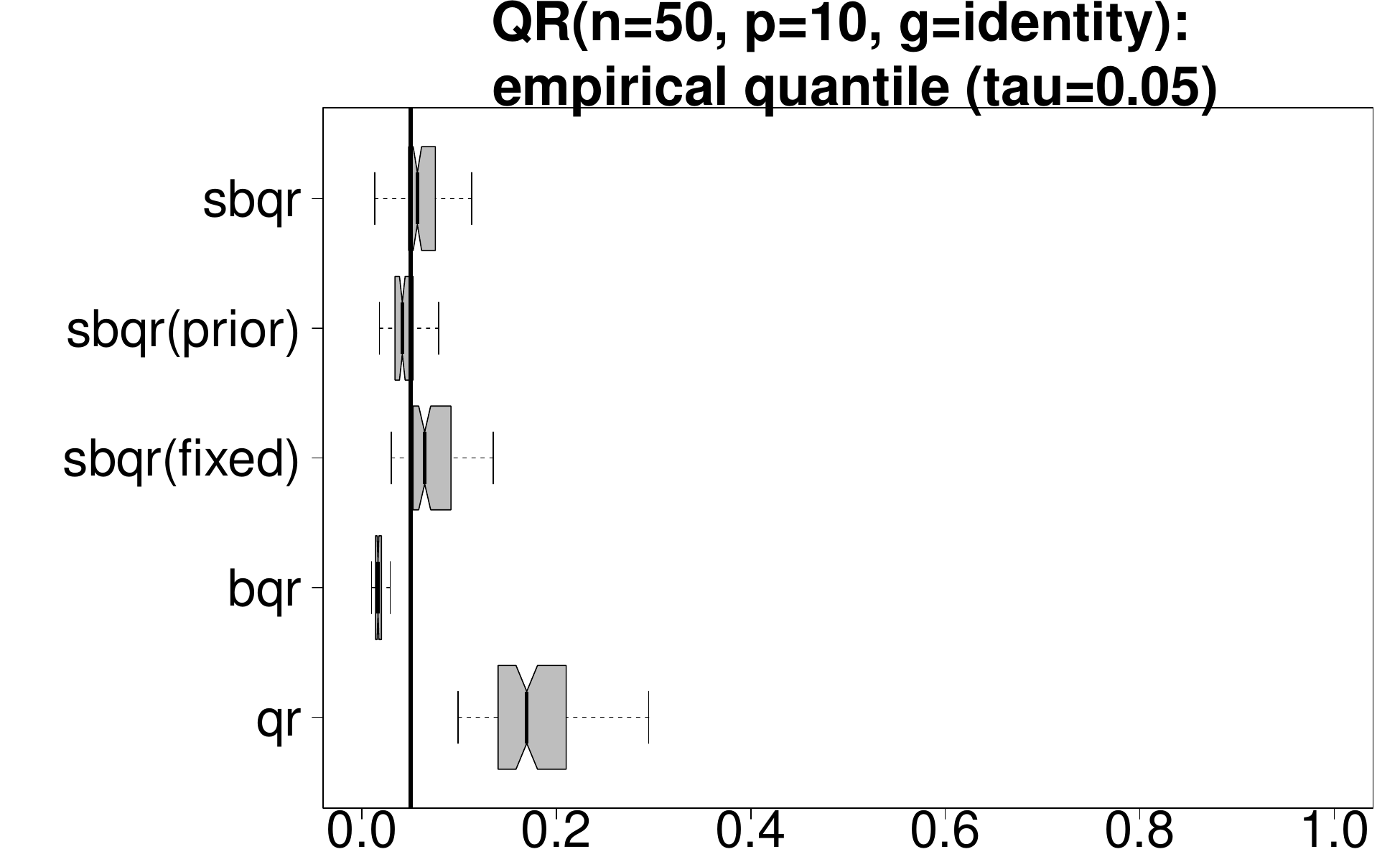}
\includegraphics[width=.32\textwidth]{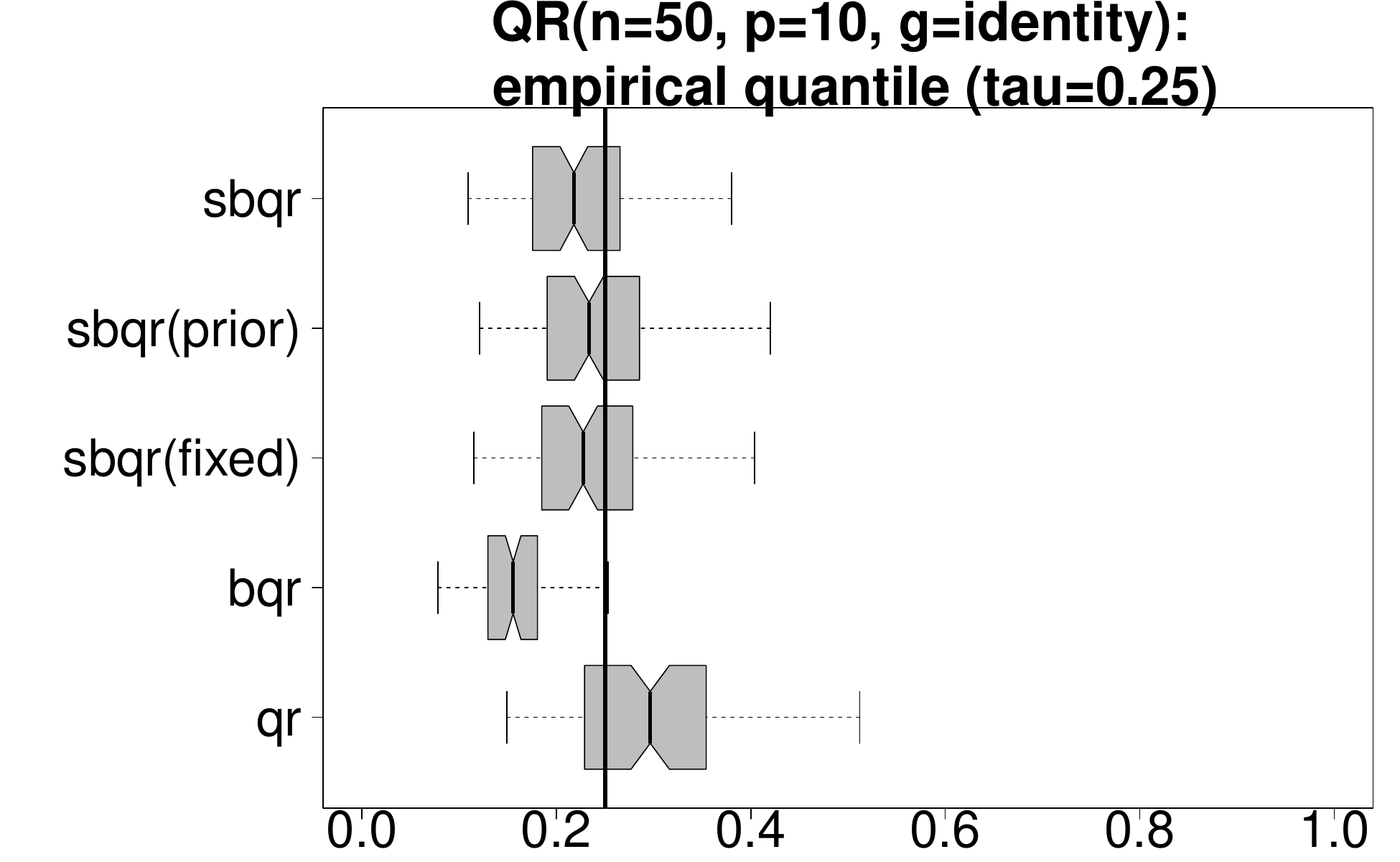}
\includegraphics[width=.32\textwidth]{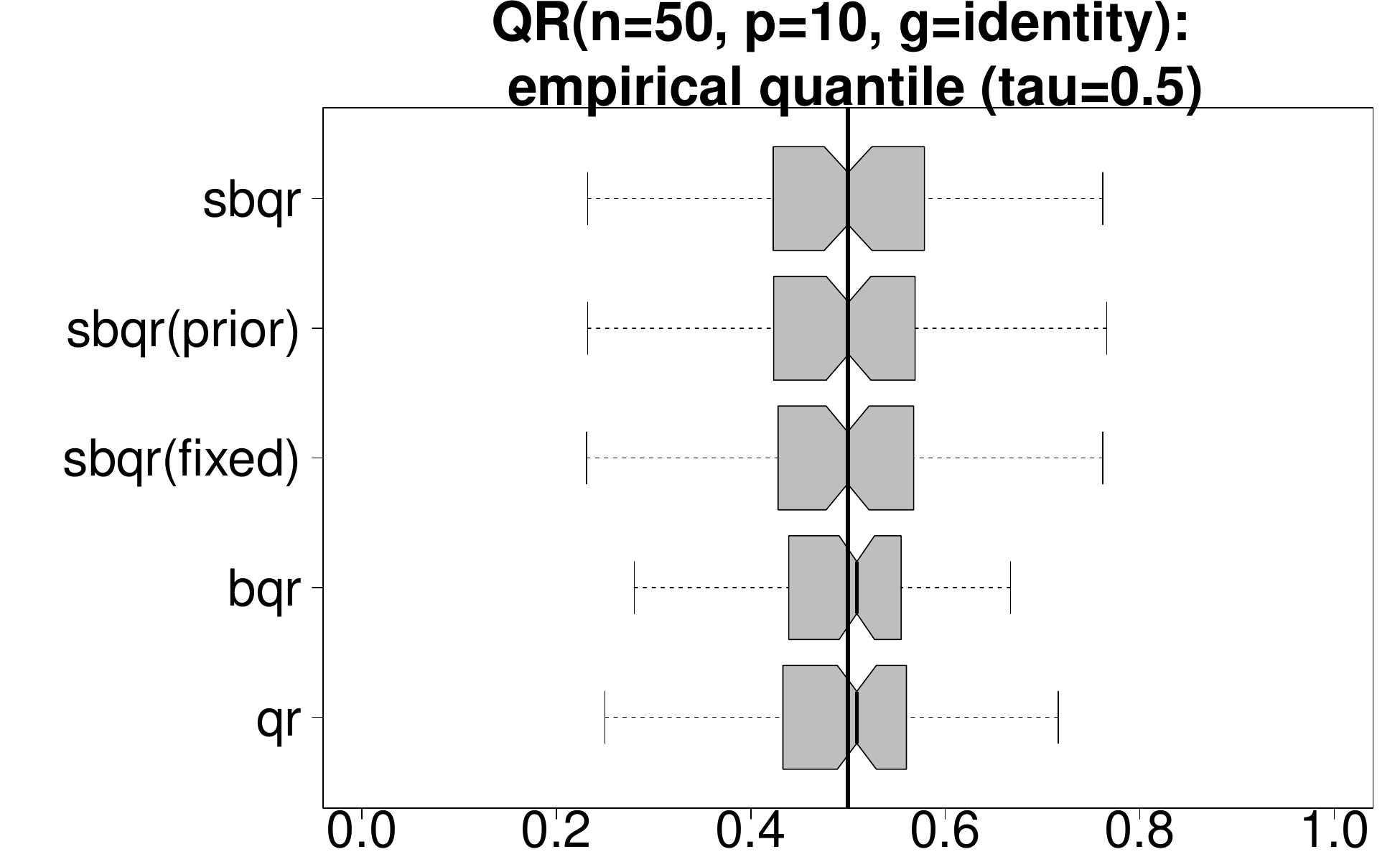}
\includegraphics[width=.32\textwidth]{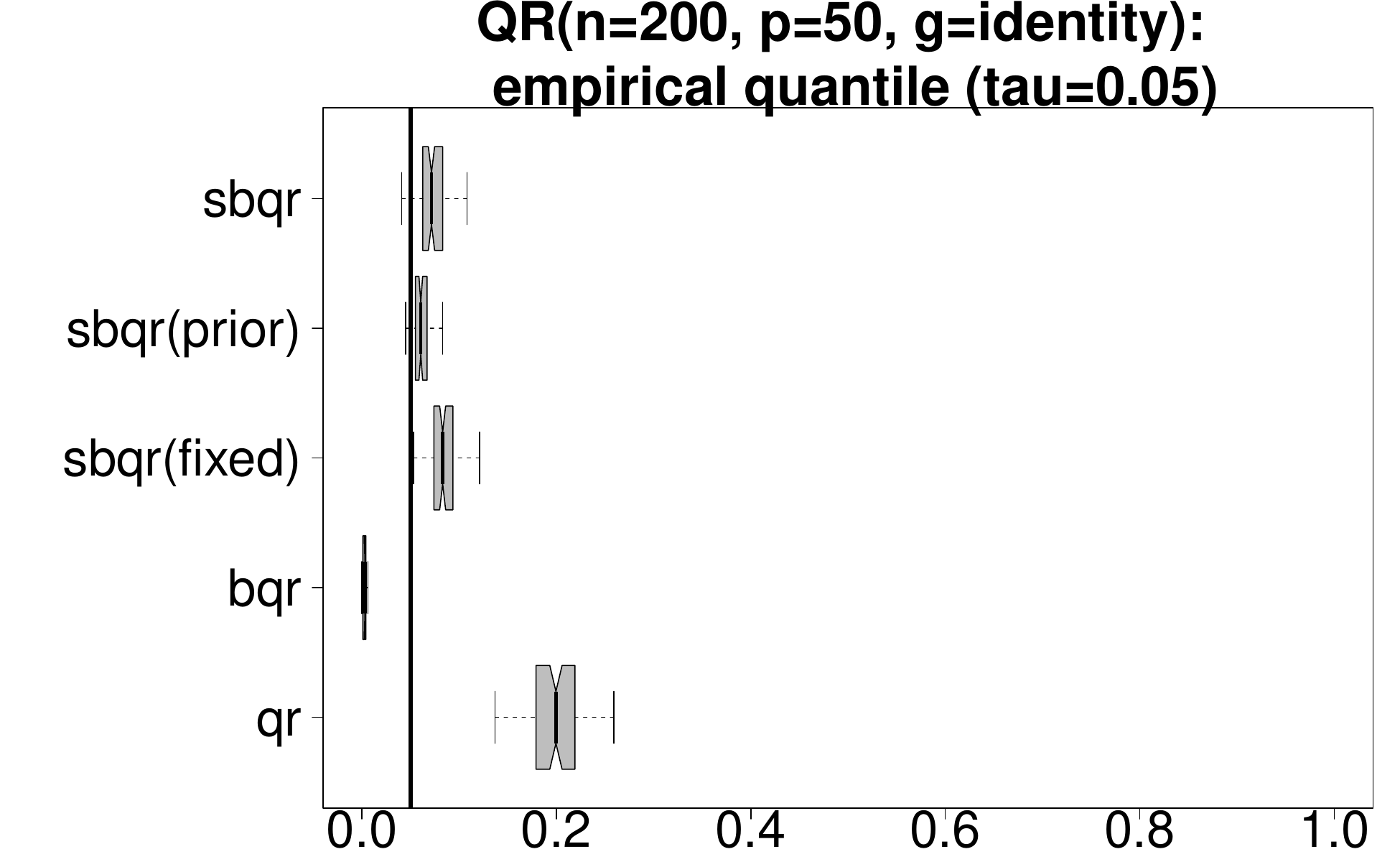}
\includegraphics[width=.32\textwidth]{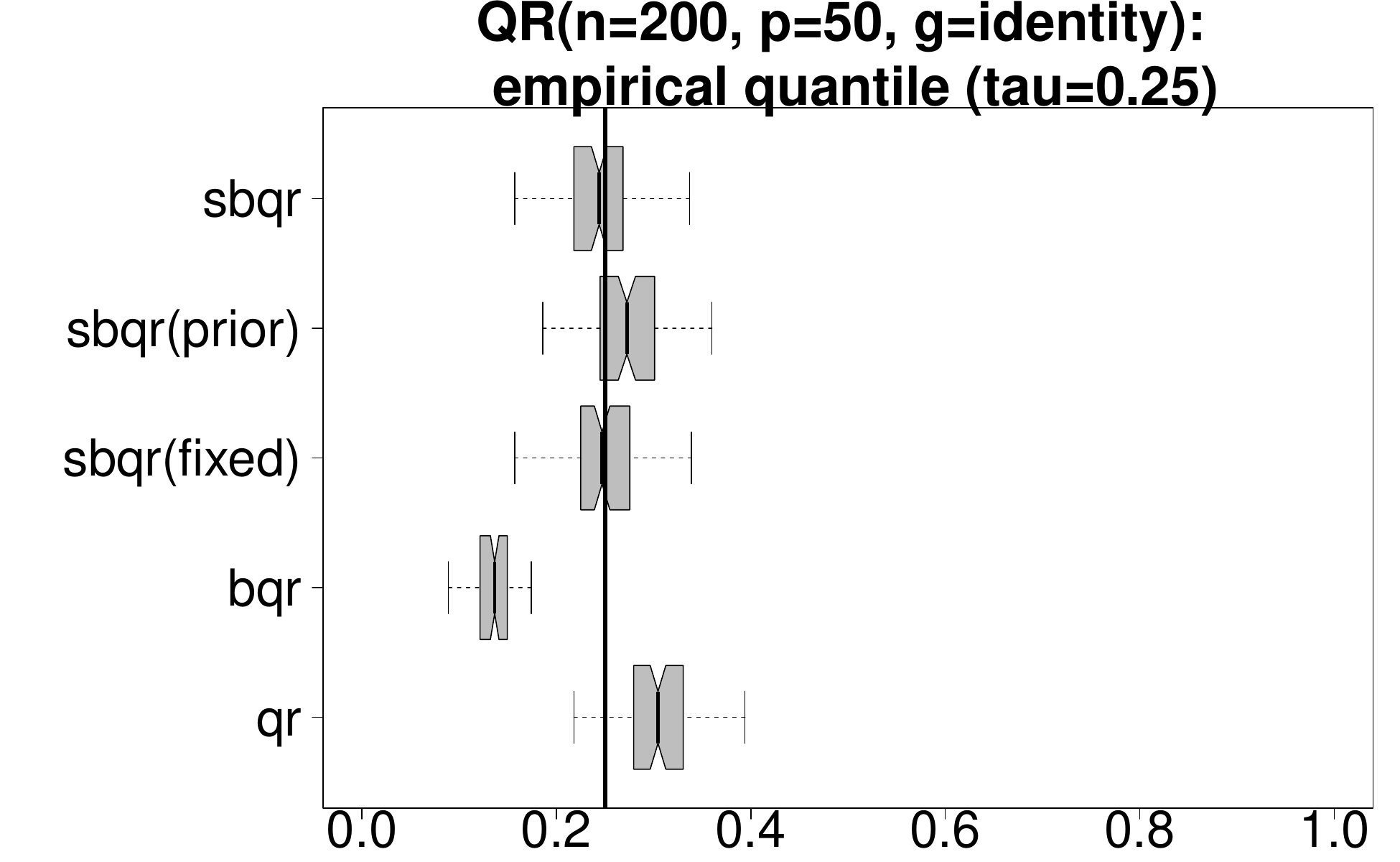}
\includegraphics[width=.32\textwidth]{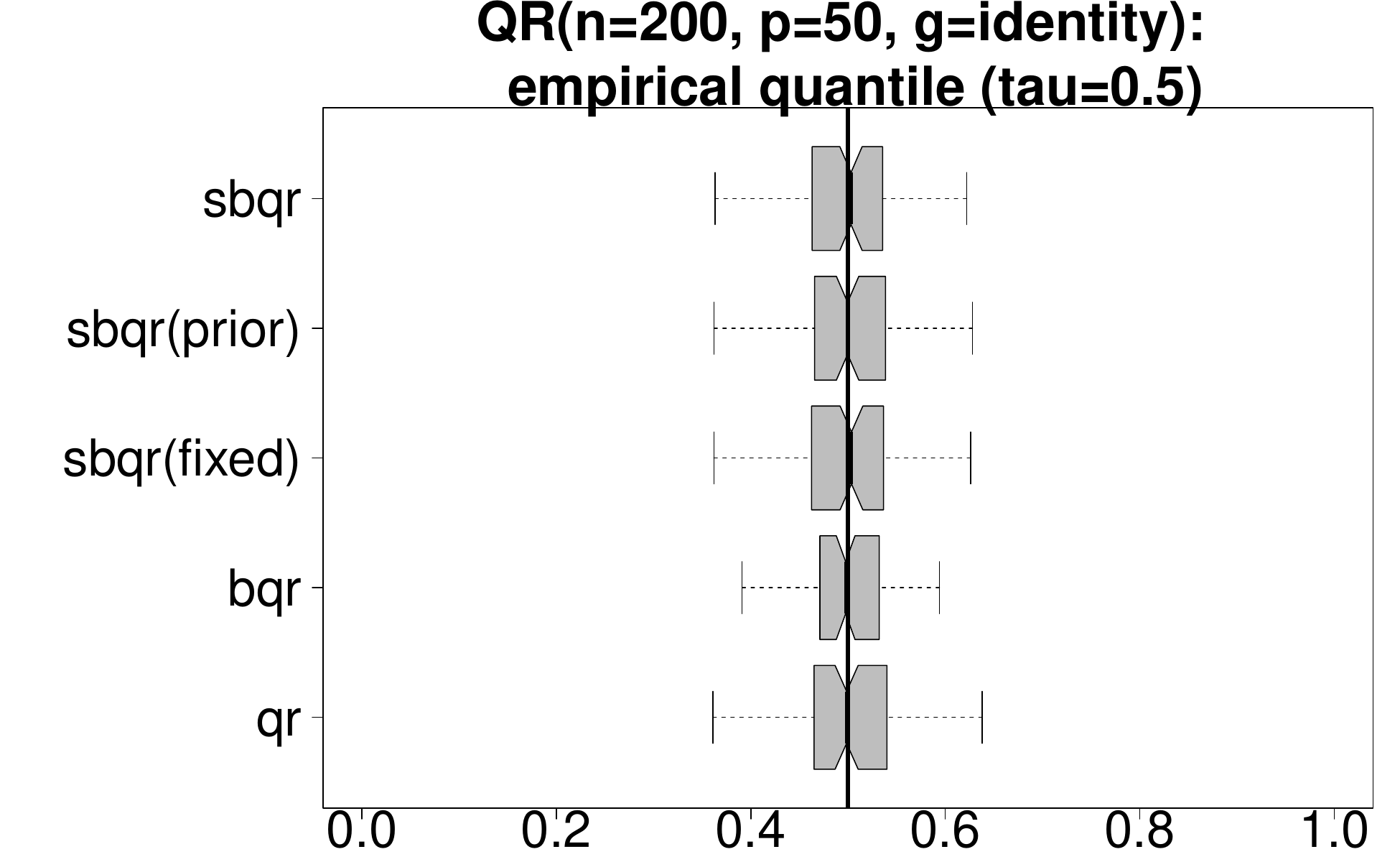}
 \caption{Proportion of testing data below the estimated $\tau$th quantile (vertical  line at $\tau$) across 100 simulations.  Each method is well-calibrated for the median $(\tau = 0.5)$, but the proposed semiparametric estimators are significantly more accurate  when $\tau$ is small (or large; not shown).
\label{fig:sims-qr}}
\end{figure}

We evaluate the quantile estimates by computing the proportion of testing data points that are below the estimated $\tau$th quantile (Figure~\ref{fig:sims-qr}).  For a well-calibrated  quantile estimate, this quantity should be close to $\tau$. Although all methods are well-calibrated for the median $(\tau = 0.5)$, the existing frequentist (\texttt{qr}) and Bayesian (\texttt{bqr}) estimates become poorly calibrated  as $\tau$ decreases (or increases; not shown). By comparison, the proposed semiparametric methods maintain   calibration across all values of $\tau$, especially for the  fully Bayesian  implementations. Again, we see little difference between the data-driven approximation (\texttt{sbqr}) and the  prior approximation (\texttt{sbqr(prior)}) central to \eqref{Fzx}, which highlights the  robustness of Algorithm~\ref{alg:Fz}.

We emphasize the value of the transformation for improving Bayesian model adequacy.  Specifically, we compute continuous ranked probability scores for the posterior predictive draws  $\tilde y(X^{test})$ from each Bayesian model, and average those across all simulations (Table~\ref{tab:rps}). These scores provide a comprehensive assessment of the out-of-sample posterior predictive distributions. Compared to the standard Bayesian quantile regression model (\texttt{bqr}), the proposed semiparametric modification offers massive improvements in predictive distributional  accuracy. In particular, \texttt{bqr} is highly inaccurate for smaller $\tau$, while the \texttt{sbqr} methods are  robust across $\tau$. Thus, the semiparametric approach alleviates concerns about the inadequacy of an ALD  model and produces more accurate quantile estimates. 


\begin{table}[h]
 \centering
\begin{tabular}{ c |  c | c  | c  | c }
 Quantile  & \texttt{bqr} & \texttt{sbqr(fixed)} & \texttt{sbqr(prior)} & \texttt{sbqr} \\ \hline
    $\tau  = 0.05$ & 7.55 & 0.50 & 0.49 &  0.50 \\ 
    $\tau  = 0.25$ & 1.07 & 0.41 & 0.41 & 0.41 \\
    $\tau  = 0.50$ & 0.66 & 0.40 & 0.40 & 0.40  \\
\end{tabular}
\caption{Continuous ranked probability scores on testing data for Bayesian quantile regression with $n=50,  p=10$;  results are  similar for $n=200,p=50$. The proposed semiparametric Bayesian quantile models offer massive improvements in predictive distributional accuracy.
\label{tab:rps}}
\end{table}

\begin{figure}[h]
\centering
\includegraphics[width=.32\textwidth]{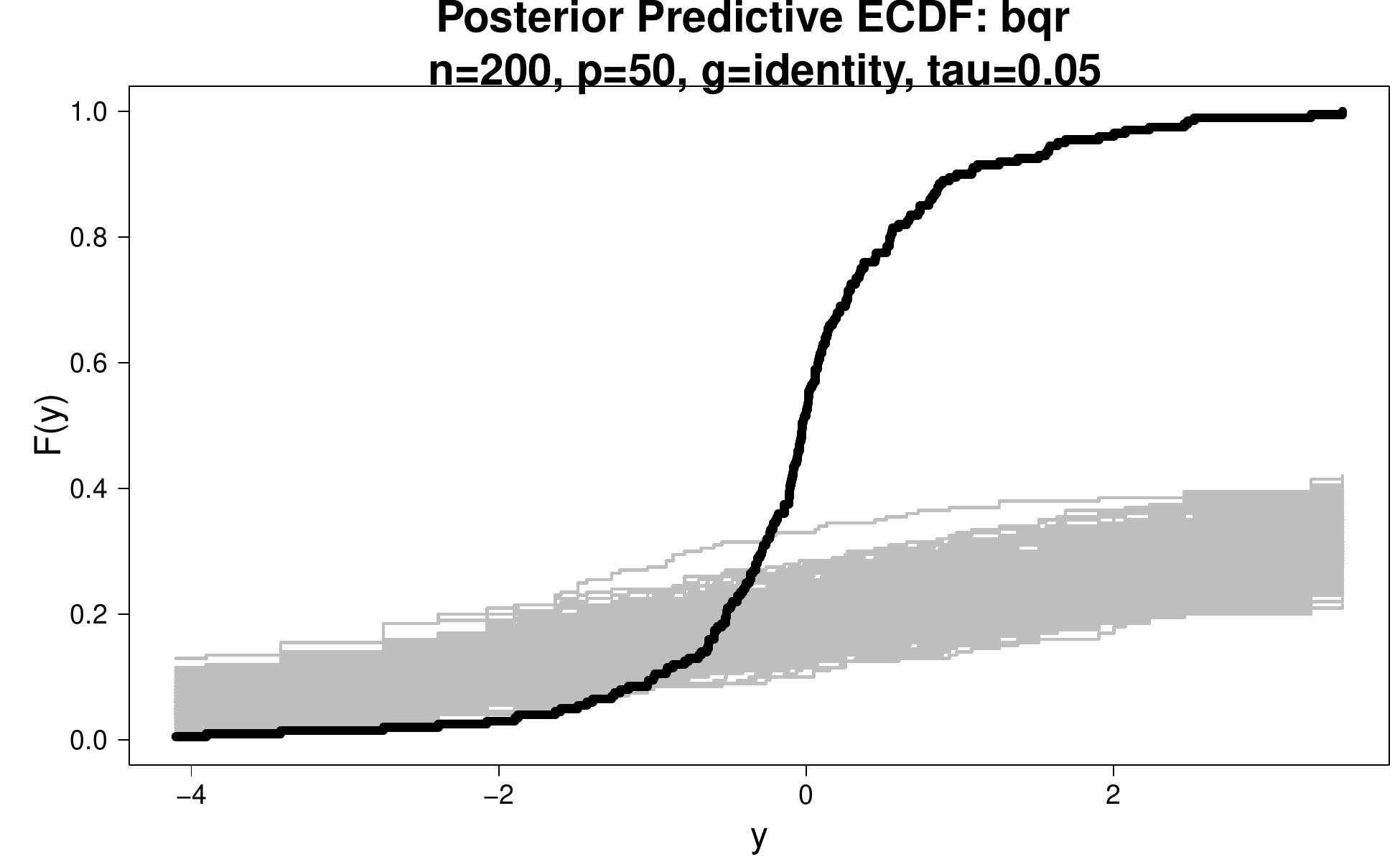}
\includegraphics[width=.32\textwidth]{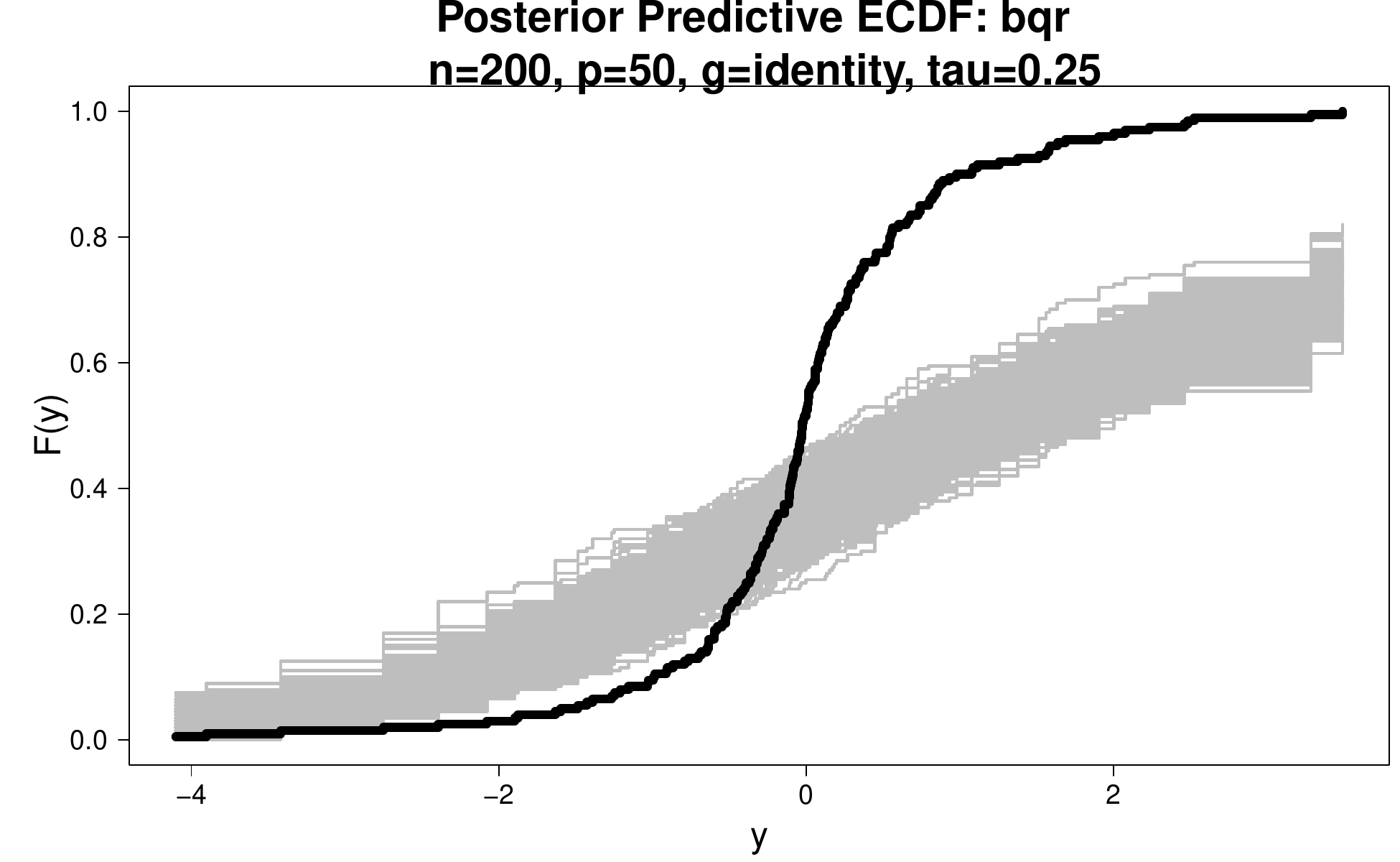}
\includegraphics[width=.32\textwidth]{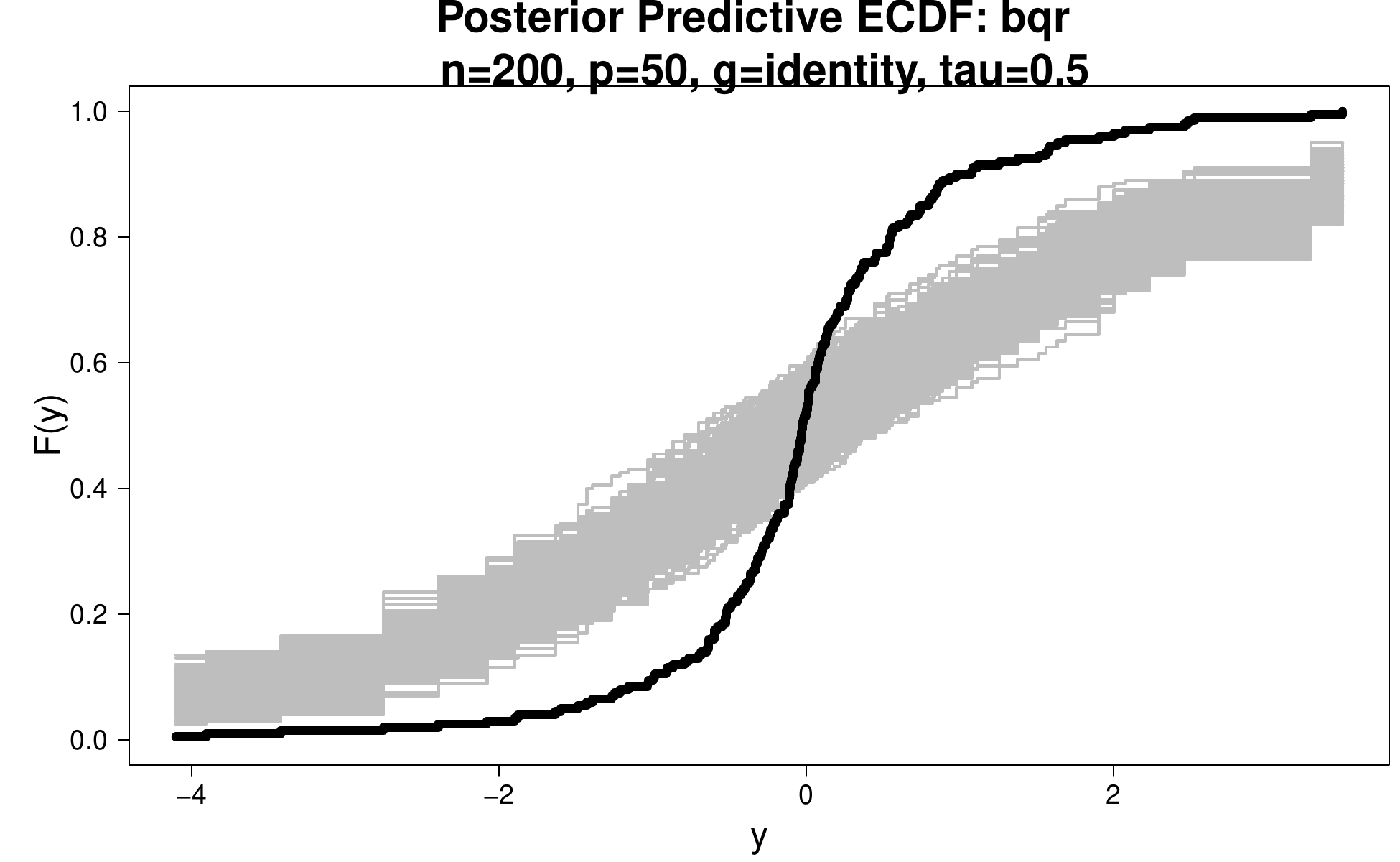}
\includegraphics[width=.32\textwidth]{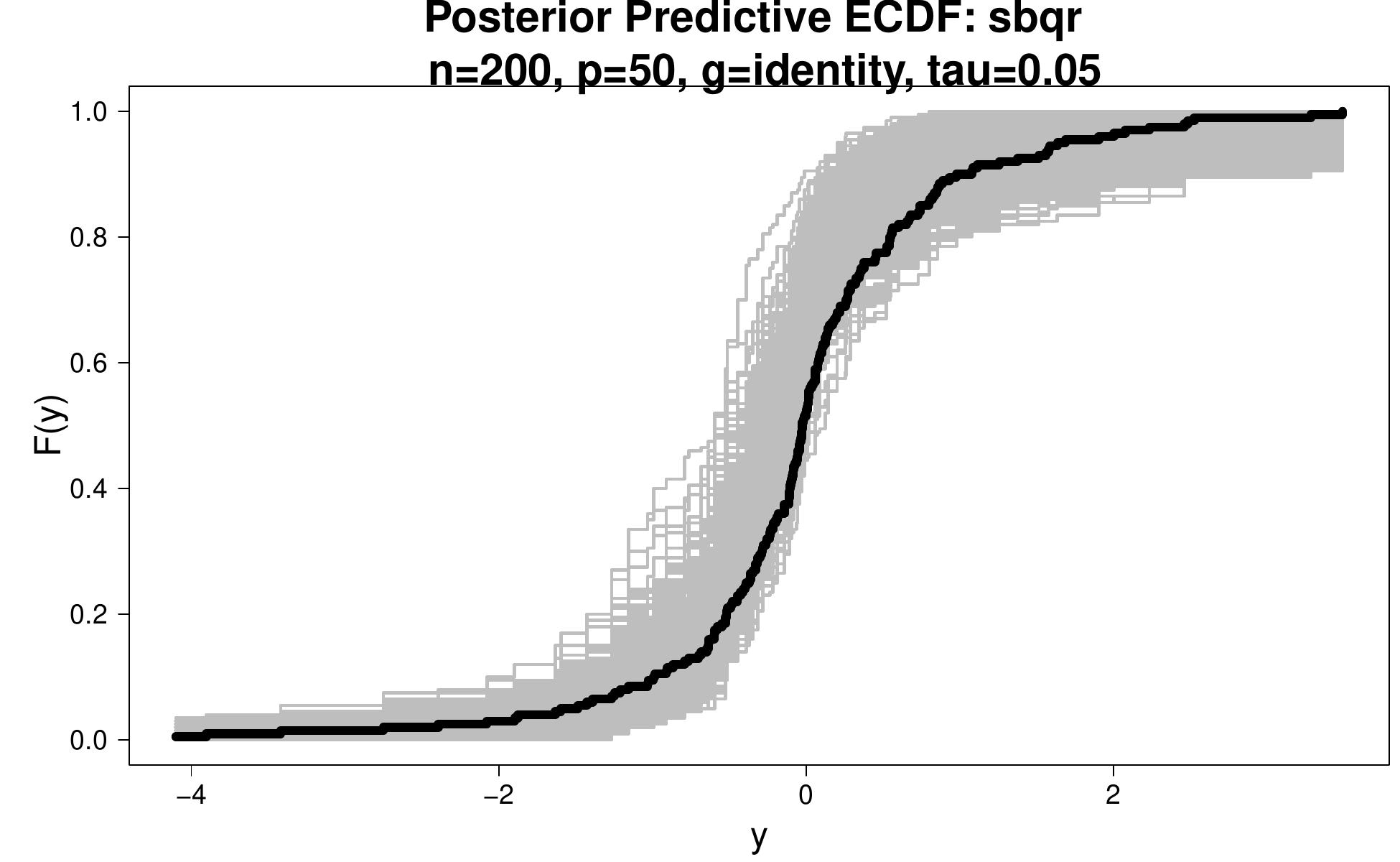}
\includegraphics[width=.32\textwidth]{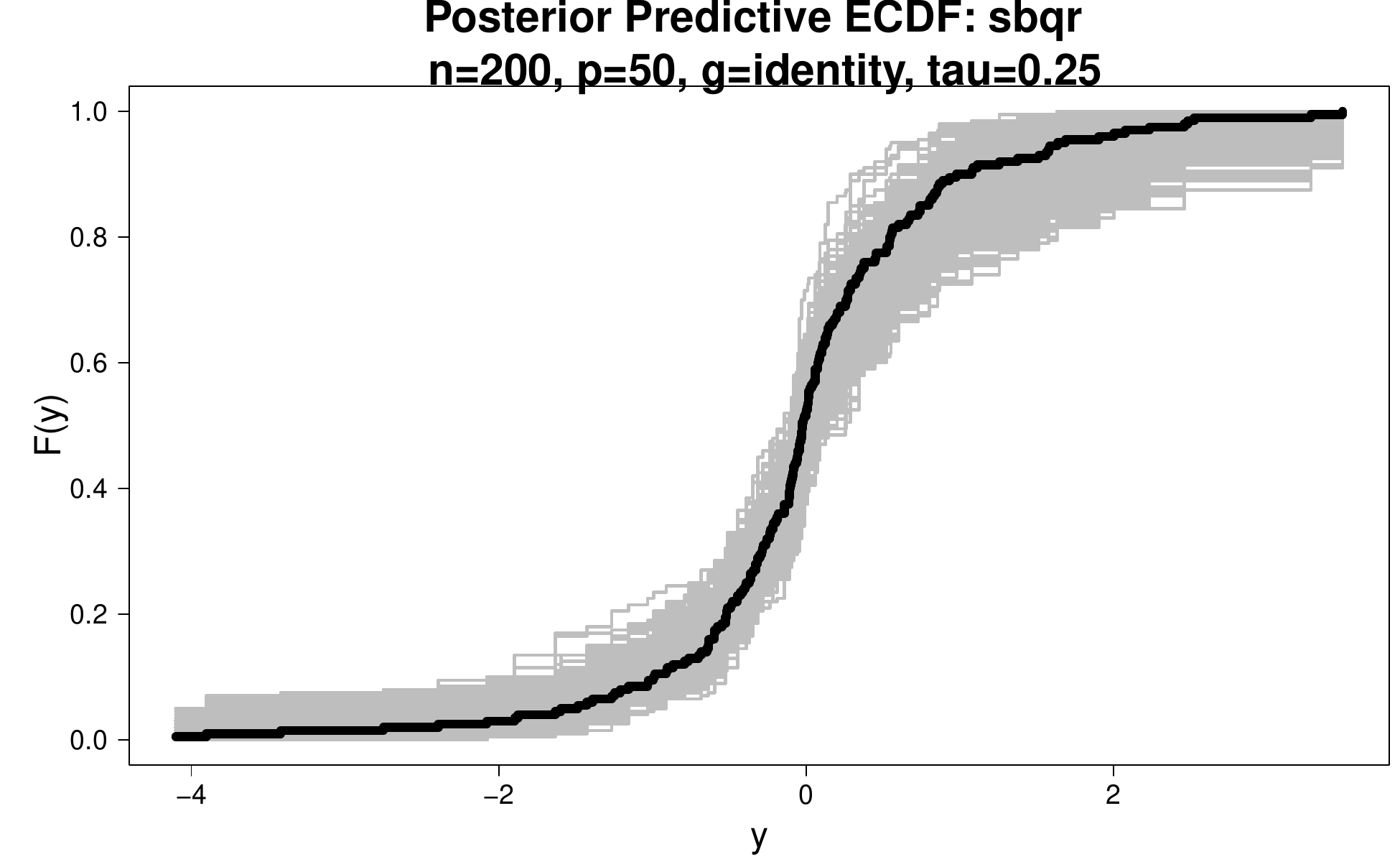}
\includegraphics[width=.32\textwidth]{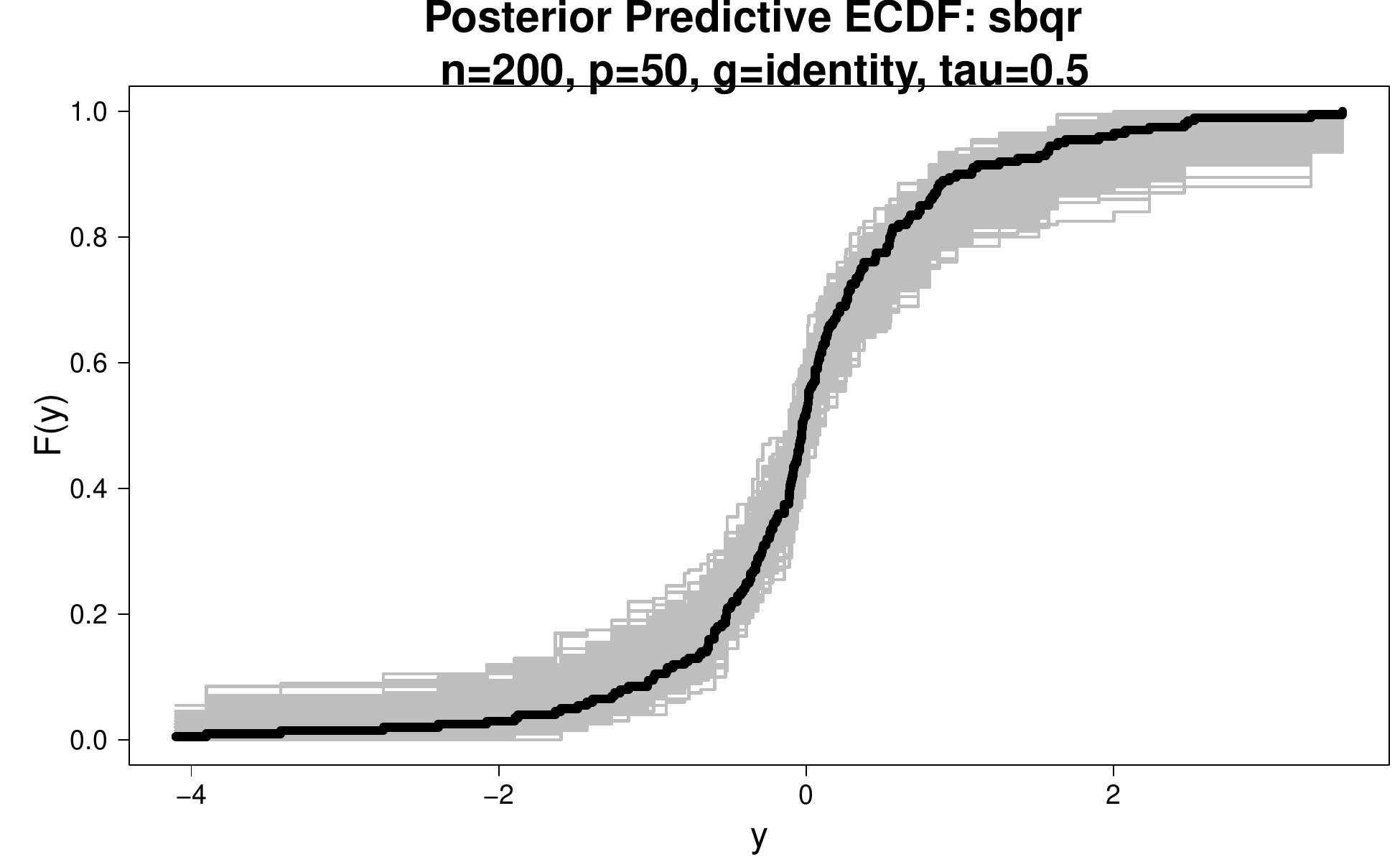}
 \caption{Posterior predictive diagnostics for Bayesian quantile regression based on the empirical cumulative distribution function for each $\tau \in \{0.05, 0.25, 0.5\}$. The posterior predictive draws (gray lines) are faithful to the observed data (black lines) for the proposed approach (bottom), while traditional Bayesian quantile regression (top) is highly inadequate. 
\label{fig:ecdf-qr}}
\end{figure}

These important conclusions are confirmed visually by examining posterior predictive diagnostics for a single simulated dataset (Figure~\ref{fig:ecdf-qr}). We compute the empirical cumulative distribution function for the observed data and for each posterior predictive draw under the Bayesian quantile regression model for each $\tau \in \{0.05, 0.25, 0.5\}$. Traditional Bayesian quantile regression based on the ALD is clearly inadequate as a data-generating process for all values of $\tau$. By comparison, the proposed semiparametric alternative \texttt{sbqr} (and \texttt{sbqr(prior)}; not shown) completely corrects these inadequacies to deliver a model that both  infers the target  quantiles accurately (Figure~\ref{fig:sims-qr}) and is globally faithful to the observed data (Table~\ref{tab:rps} and Figure~\ref{fig:ecdf-qr}).

\subsection{Semiparametric Gaussian processes for Lidar data}\label{sec-lidar}
\begin{figure}[h]
\centering
\begin{minipage}[b]{0.59\textwidth}
\includegraphics[width=1\textwidth]{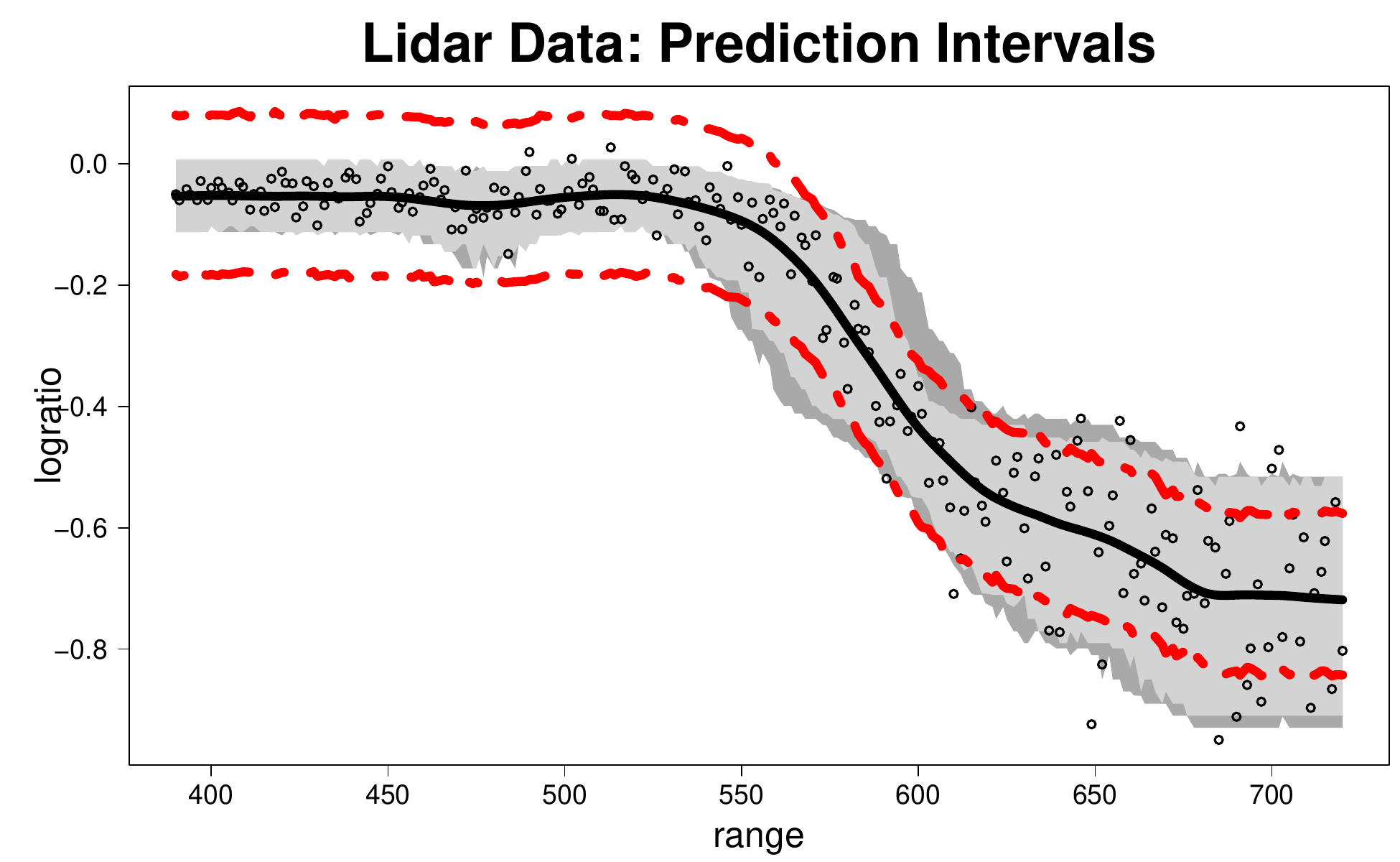}
\end{minipage}
\begin{minipage}[b]{0.4\textwidth}
\includegraphics[width=.7\textwidth]{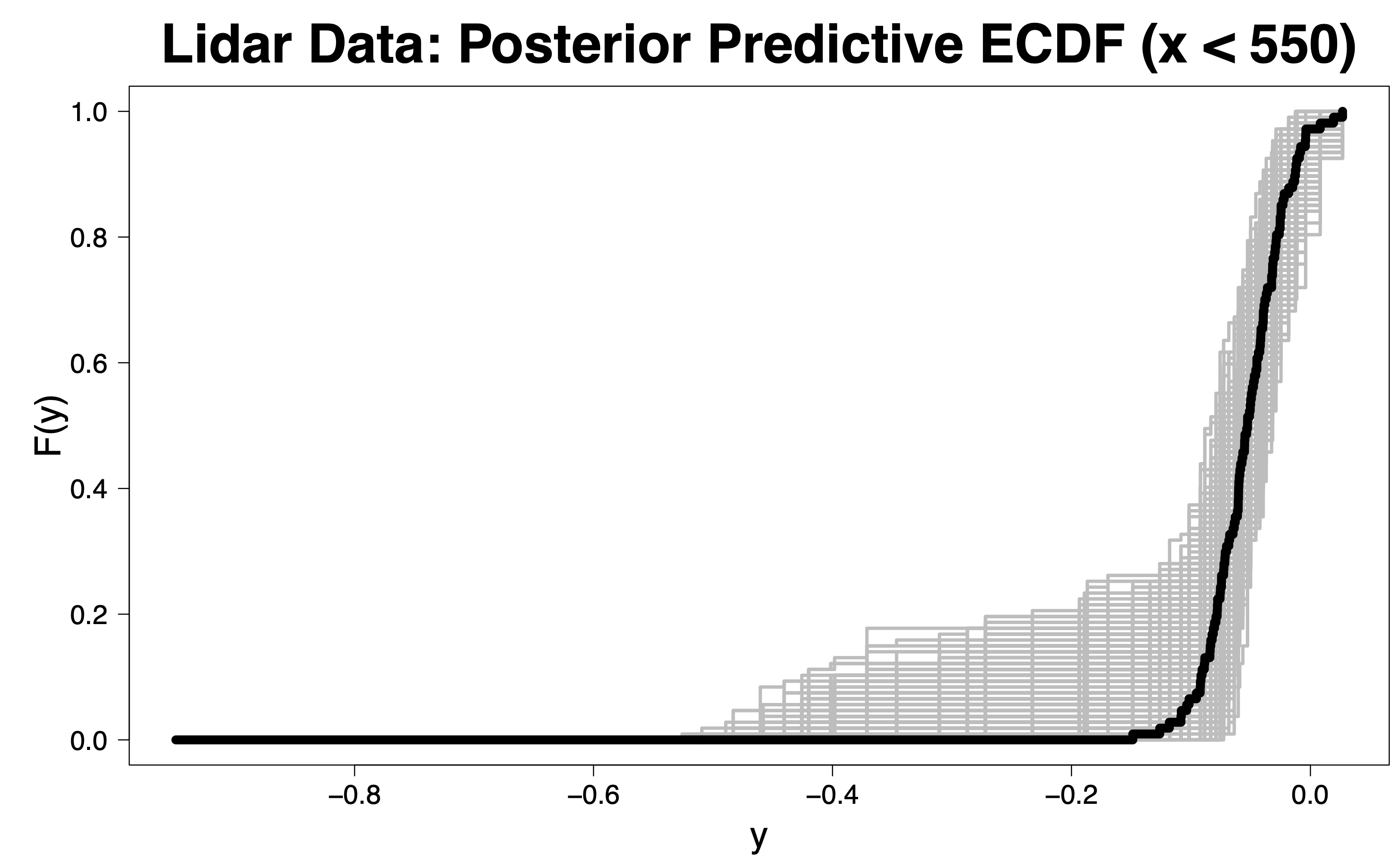}

\includegraphics[width=.7\textwidth]{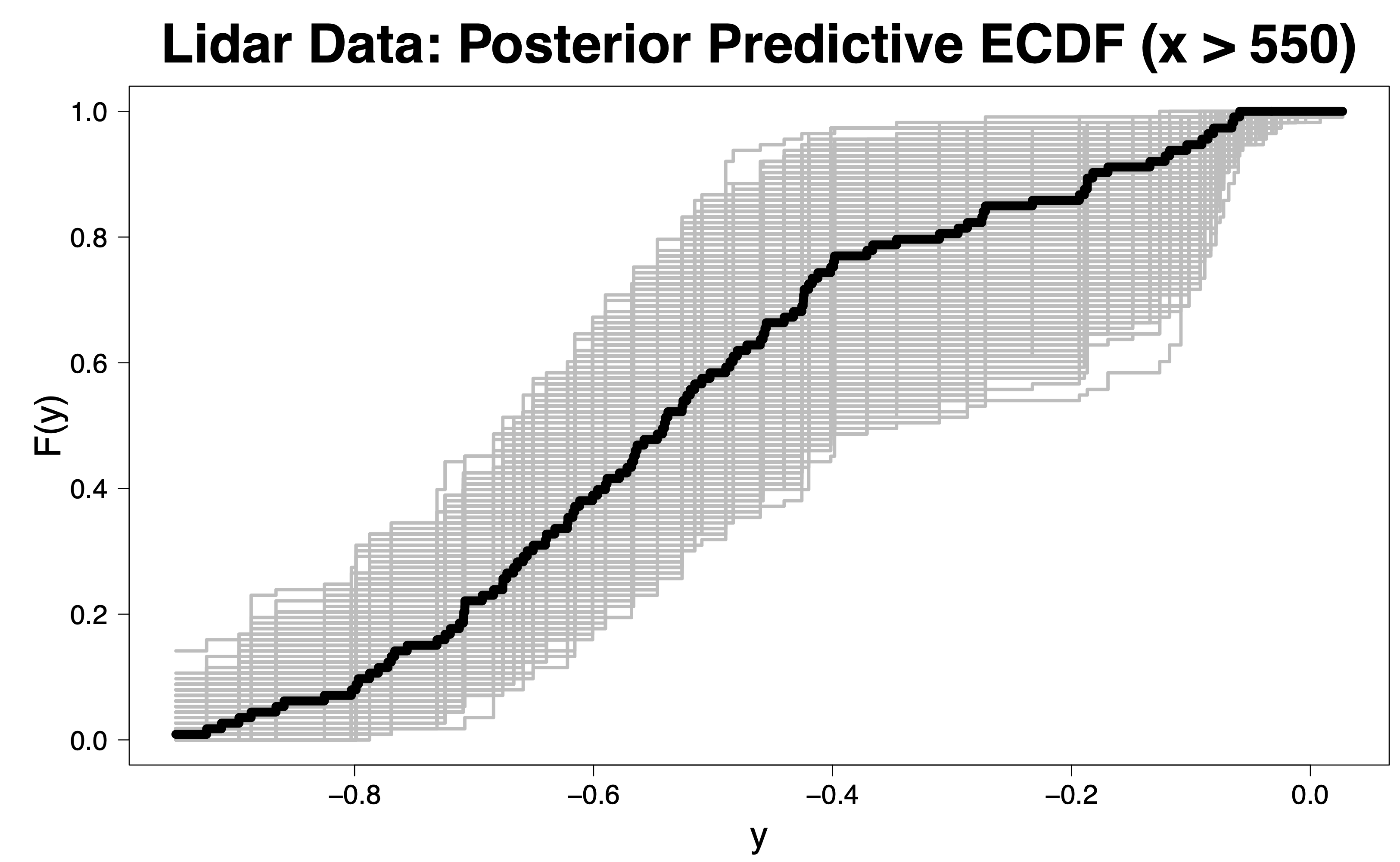}
\end{minipage}
\caption{Left:  90\% prediction intervals and median (black line) for the Lidar data. The proposed semiparametric Gaussian process model (dark gray) and approximate version (light gray) capture the heteroskedasticity in the data,  while the  Box-Cox version  (dashed red)  does not. Right: Posterior predictive diagnostics based on the marginal empirical cumulative distribution function  restricted version for smaller (top) and larger (bottom)  $x$ values. The posterior predictive draws (gray lines) from \texttt{sbgp} are faithful to the observed  data (black lines).
\label{fig:lidar-in}}
\end{figure}

We apply the semiparametric Bayesian Gaussian process model to the Lidar data from \cite{ruppert2003semiparametric}. These data are a canonical example of a nonlinear and heteroskedastic curve-fitting problem (Figure~\ref{fig:lidar-in}). Instead of augmenting a Gaussian process model with a variance model---which requires additional model specification, positivity constraints, and more demanding computations---we simply apply the proposed semiparametric Bayesian Gaussian process (\texttt{sbgp}) approach from Section~\ref{sec-sbgp}. The latent Gaussian process $f_\theta \sim \mathcal{GP}(m_\theta, K_\theta)$ features an unknown constant for the mean function $m_\theta$ and an isotropic Mat\'ern covariance function $K_\theta$ with unknown variance, range, and smoothness parameters; these unknowns constitute $\theta$. Computations of the Gaussian process point estimates, predictions, and covariances are done using the 
 \texttt{GpGp} package in \texttt{R} 
\citep{Katzfuss2021}.

First, we assess the \texttt{sbgp} fit to the full dataset ($n=221$). Figure~\ref{fig:lidar-in} (left) presents the fitted curve and 90\% pointwise prediction intervals. The \texttt{sbgp} model is capable of smoothly capturing the trend \emph{and} the heteroskedasticity in the data---which is not explicitly modeled. More thorough posterior predictive diagnostics (Figure~\ref{fig:lidar-in}, right) confirm the adequacy of the model. Specifically, we compute the empirical cumulative distribution on the data and on each simulated predictive dataset, in both cases restricted  to smaller ($x < 500$) and larger ($x > 500$) covariate values. Despite the notable differences in the distributions, the proposed \texttt{sbgp}  is faithful to the data.

For comparison, we include an approximate version that fixes the transformation at   $\hat g$ 
(\texttt{sbgp(fixed)}), and consider more standard Gaussian process models that omit the transformation (\texttt{gp}) or apply an unknown Box-Cox transformation (\texttt{gp(box-cox)}) using the same prior and sampler for $\lambda$ as in Section~\ref{sec-lm}. Figure~\ref{fig:lidar-in} shows that \texttt{sbgp(fixed)} naturally produces narrower prediction intervals, but more importantly, that \texttt{gp(box-cox)} is unable to capture the heteroskedasticity in the data. Thus, unlike the proposed nonparametric Bayesian transformation, the parametric Box-Cox transformation is inadequate for this key distributional feature.

We proceed with more formal evaluations based on 100 random training/testing  splits of the data (80\% training). 
Table~\ref{tab:lidar-miw} presents the average interval widths and empirical coverage for 95\%, 90\%, and 80\% out-of-sample prediction intervals. Remarkably, \texttt{sbgp} delivers the exactly correct nominal coverage, often with narrower intervals than the competing \texttt{gp} and \texttt{gp(box-cox)} models. The intervals from the approximate version are narrower and typically close to nominal coverage. The proposed methods are both calibrated and sharp. This is confirmed using out-of-sample ranked probability scores (not shown), which show statistically significant improvements for \texttt{sbgp(fixed)} and \texttt{sbgp} relative to the Gaussian process competitors. 

\begin{table}[h]
\centering
\begin{tabular}{c | c | c | c | c}
 Nominal coverage & \texttt{gp} & \texttt{gp(box-cox)} & \texttt{sbgp(fixed}) & \texttt{sbgp} \\ 
  \hline
95\% & 0.311 (92\%) & 0.313 (92\%)  & 0.277 (93\%) & 0.326 (95\%) \\ 
90\% & 0.261 (90\%) & 0.263 (90\%)  & 0.224 (89\%) & 0.256 (90\%) \\ 
80\% & 0.204 (86\%) & 0.205 (86\%) & 0.171 (80\%) & 0.195 (81\%) 
\end{tabular}
\caption{Average interval width and empirical coverage (parenthetical) for  prediction intervals for Lidar data across 100 random training/testing splits. The proposed 
approaches are well-calibrated (correct coverage) and sharper (narrower intervals) than competing methods.
\label{tab:lidar-miw}}
\end{table}

Finally, we modified the approximate approach from Section~\ref{sec-sbgp} to include posterior sampling of $f_{\hat \theta}$, which accounts for the uncertainty in the regression function (but not the hyperparameters $\theta$). The results were visually indistinguishable from Figure~\ref{fig:lidar-in}, but the computing cost increased substantially: for the full dataset, \texttt{sbgp} required only about 4 seconds per 1000 Monte Carlo samples, 
while the augmented posterior sampler needed about 111 seconds. These  discrepancies increase with $n$. In aggregate, our results suggest that \texttt{sbgp} successfully combines the efficiency of point optimization with the uncertainty quantification from the Bayesian nonparametric model for $g$ to deliver fast, calibrated, and sharp posterior predictive inference.  



\section{Theory}\label{sec-theory}

An advantage of our modeling and algorithmic framework is that it  enables direct asymptotic analysis. We consider generic models for $P_X$ and $P_Y$ within model \eqref{trans}--\eqref{mod} and show that our joint posterior for $(g, \theta)$ under Algorithm~\ref{alg:joint} is consistent under conditions on $P_Y$, $P_X$, and $P_{Z\mid \theta, X}$. Here, $P_{Z\mid \theta, X}$ is permitted to be more general than \eqref{mod}, but we require finite-dimensional parameters (e.g., Sections~\ref{sec-sblm}~and~\ref{sec-sbqr}); modifications for Gaussian processes (Section~\ref{sec-sbgp}) are in the supplementary material. Importantly, these results verify the asymptotic validity of (i) the surrogate likelihood, (ii) the approximation \eqref{Fzx}, and (iii) the location-scale robustness adjustment (Section~\ref{sec-approx}).  

Let $\M(\Y)$ denote the space of monotone increasing functions mapping $ \Y$ to $\R$ and let $\mathcal{T}$ be the topology of pointwise convergence. Let  $F_{X,0}$ and $F_{Y,0}$ denote the true distribution functions of $X$ and $Y$, respectively. Finally, let $\tilde g$ be the restriction of $g$ to $\tilde \Y = \{y \in \Y: 0 < F_{Y,0}(y) < 1\}$ and $\tilde g_0$ defined similarly for the true transformation $g_0$. 

First, we establish posterior consistency for $p(\tilde g \mid \D_n)$ at  $\tilde g_0$.
 \begin{theorem}
    \label{thm:cons-g}
Suppose that the true data-generating process $P_{g_0, \theta_0}$ is identified by the parameters $  g_0\in  (\M(\Y), \mathcal{T})$ and $\theta_0  \in \Theta$ under model \eqref{trans}--\eqref{mod} with $\Theta \subseteq \R^d$ open. Under the assumptions 
\begin{enumerate}[(\ref{thm:cons-g}.1)]
\vspace{-2mm}\item $F_{Z \mid \theta, X}(t)$ is continuous in $X$, $t$, and in $\theta$ for an open neighborhood of $\theta_0$ invariant to $(t, X)$;
\vspace{-2mm}\item the posterior approximation $\hat p( \theta \mid \D_n)$ is (strongly) consistent at $\theta_0$; and  
\vspace{-2mm} \item the marginal models $F_{X \mid \D_n}, F_{Y \mid \D_n}$ are (strongly) consistent at $F_{X,0}, F_{Y,0}$,
\vspace{-2mm} \end{enumerate}
then the posterior distribution $p(\tilde g \mid \D_n)$ is (strongly) consistent at $\tilde g_0$ under (i) $\Tau$ and (ii)  the $L_\infty$-topology on any bounded subset of $\tilde \Y$. 
\end{theorem}
Importantly, this posterior  $p(\tilde g \mid \D_n)$ refers to the one targeted by Algorithm~\ref{alg:joint},  which uses (i) the surrogate likelihood and (ii) the approximation $\hat p(\theta \mid \D_n)$ in \eqref{Fzx}. 
The regression models from Sections~\ref{sec-sblm}~and~\ref{sec-sbqr} satisfy (\ref{thm:cons-g}.1), while the BB models for  $P_X$ and $P_Y$ are (weakly) consistent for  the margins and satisfy (\ref{thm:cons-g}.3). 
Empirical support for Theorem~\ref{thm:cons-g} is provided in Figure~\ref{fig:sims-trans} and the supplementary material. When $X$ is nonrandom, we define $F_{X,0}$ to be the empirical distribution of the fixed covariate values $\{x_i\}_{i=1}^n$. The accompanying modification of Algorithm~\ref{alg:Fz}  with $\alpha_i^x = 1/n$ implies  $F_{X \mid \D_n}=F_{X,0}$, which relaxes (\ref{thm:cons-g}.3) and otherwise maintains Theorem~\ref{thm:cons-g}.

The requirement (\ref{thm:cons-g}.2) admits many choices of $\hat p(\theta \mid \D_n)$. Perhaps the simplest option is a point mass at some consistent estimator of $\theta_0$, such as using rank-based estimators (\citealp{Horowitz2012}; see the supplementary material). 
\begin{corollary}\label{cor-freq-const}
Let $\hat  p(\theta \mid \D_n) = \delta_{\tilde \theta}$, where $\tilde \theta \overset{p}{\to} \theta_0$ is a consistent point estimator. Under the conditions of Theorem~\ref{thm:cons-g}, 
 $p(\tilde g \mid \D_n)$ is weakly consistent at $\tilde g_0$ under $\Tau$. 
\end{corollary}
To strengthen Theorem~\ref{thm:cons-g}, we consider special cases of the support $\Y$.  The restriction to $\tilde \Y$ ensures that $g$ is finite, but the limiting cases are  well-defined: we may set $g(t) = -\infty$ for any $t$ such that $F_{Y,0}(t) = 0$ and similarly $g(t) = \infty$ whenever $F_{Y,0}(t)=1$. No such consideration is needed when $\Y$ is unbounded. 
\begin{corollary}\label{cor-unbounded}
Suppose $\Y$ is unbounded. Under the conditions of Theorem~\ref{thm:cons-g},  $p(g \mid \D_n)$ is (strongly) consistent at $g_0$ under (i) $\Tau$ and (ii) the $L_\infty$-topology on any bounded subset of $\tilde \Y$.  
\end{corollary}
When $\Y$ is compact, we can strengthen this result to uniform posterior consistency of $p(g \mid \D_n)$ with no restrictions. Uniform posterior consistency ensures that the posterior converges to the true parameter at the same rate in all regions of the domain. 
\begin{corollary}
\label{cor:con-g-compact}
Suppose $\Y$ is compact. Under the conditions of Theorem~\ref{thm:cons-g},  $p( g \mid \D_n)$ is weakly consistent at $g_0$ under the $L_\infty$-topology.
\end{corollary}
%


Building upon Theorem~\ref{thm:cons-g}, we  establish joint posterior consistency of $(g, \theta)$ by showing  consistency of the conditional posterior $p(\theta \mid \D_n, g)$ with fixed transformation $g$.   
For robustness, we provide sufficient conditions for strong posterior consistency of $p(\theta \mid \D_n, g)$ \emph{without} assuming the correctness of the model for $\theta$. Instead, we target the parameter that minimizes Kullback–Leibler (KL) divergence from an arbitrary data-generating process to the assumed model.  
\begin{theorem}\label{thm:theta-con} 
Let $P_0$ be the true data-generating process and $P_{g,\theta}$ the data-generating model induced by \eqref{trans}--\eqref{mod} with $g\in (\M(\Y), \mathcal{T})$ and $\theta \in \Theta \subseteq \mathbb{R}^d$. Let $\Pi_\theta$ be the prior on $\theta$ and $p(Y \mid g, \theta, X)$ the likelihood of  $\theta$ at $(X,Y)$ and conditional on $g$. Under the assumptions 
\begin{enumerate}[(\ref{thm:theta-con}.1)]
\vspace{-2mm}\item there exists a unique $\theta^*(g) \in \mbox{int}(\Theta)$ such that $\theta^*(g) = \arg\min_{\theta \in \Theta} KL(P_0, P_{g, \theta})$; 
\vspace{-2mm}\item $\vert \mathbb{E}_{P_0}\log p(Y \mid g, \theta,  X) \vert <\infty$ for all $\theta \in \Theta$;
\vspace{-2mm}\item the mapping $\theta \mapsto \log p(Y \mid g, \theta, X)$ is concave almost surely $[P_0]$; and
\vspace{-2mm} \item $\Pi_\theta(\U) > 0$ for any open neighborhood $\U$ that contains $\theta^*(g)$,
\vspace{-2mm} \end{enumerate}
then $p(\theta \mid \D_n, g)$ is strongly consistent at $\theta^*(g)$ under the Euclidean topology for every fixed $g$.
\end{theorem}
The target posterior $p(\theta \mid \D_n, g) = p(\theta \mid \{x_i, g(y_i)\}_{i=1}^n)$ is equivalently the posterior distribution under \eqref{mod} using transformed data $\{x_i, g(y_i)\}_{i=1}^n$ with known $g$. Thus, some form of posterior consistency is unsurprising for many continuous regression models \eqref{mod};  the supplementary material notes conditions for the Gaussian process case \citep{choi2007posterior}.
Instead, Theorem~\ref{thm:theta-con} is valuable because (i) it establishes strong consistency for $\theta^*(g)$ under model misspecification  and (ii) it may be combined with the previous (strong) consistency results for  $p(g \mid \D_n)$ to  establish the joint posterior consistency of $(g, \theta)$. 
\begin{corollary} \label{cor-joint-con}
Suppose that the true data-generating process $P_{g_0, \theta_0}$ is uniquely identified by the parameters $  g_0\in  (\M(\Y), \mathcal{T})$ and $\theta_0  \in \Theta$ under model \eqref{trans}--\eqref{mod} with $\Theta \subseteq \R^d$ open. If the assumptions (\ref{thm:cons-g}.1)--(\ref{thm:cons-g}.3), (\ref{thm:theta-con}.1)--(\ref{thm:theta-con}.4) hold for all $g$, and if $\theta^*(g)$ is continuous at $g_0$, then the joint posterior distribution $p(\tilde{g}, \theta \mid \D_n)$ is (strongly) consistent at $(\tilde{g}_0, \theta_0)$. 
\end{corollary}
The conditions (\ref{thm:cons-g}.2)--(\ref{thm:cons-g}.3) refer to our model for $g$ under \eqref{trans-cdf} and Algorithm~\ref{alg:joint}, while (\ref{thm:cons-g}.1)  and (\ref{thm:theta-con}.1)--(\ref{thm:theta-con}.4) are regularity requirements on the model \eqref{mod} and the prior for $\theta$. 
Additional restrictions such as those in Corollaries~\ref{cor-freq-const}--\ref{cor:con-g-compact} may be applied similarly as before. We empirically confirm these asymptotics for the semiparametric linear model (Section~\ref{sec-sblm}) in the supplementary material.

Finally, we assess the proposed robustness strategy to misspecification of  $F_{Z \mid \D_n}$. When $F_{Z\mid \D_n}$ is misspecified in location or scale, accurate estimation of $g$ is impossible (Lemma~\ref{g-scale}). We consider the case in which $p(g\mid \D_n)$ converges to the wrong transformation, and specifically one that differs by a shift and scaling. The proposed strategy (Section~\ref{sec-approx}) is to replace $P_{Z \mid \theta, x}$ with $\mu + \sigma P_{Z \mid \theta, x}$ in \eqref{mod}, but \emph{only} for the conditional posterior $p(\theta \mid \D_n, g)$ in \eqref{decomp}. 
\begin{theorem} \label{thm:theta-robustness}
Let $P_0$ be the true data-generating process and $P_{g,\theta}$ the data-generating model induced by \eqref{trans}--\eqref{mod} with $g\in (\M(\Y), \mathcal{T})$ and $\theta \in \Theta \subseteq \mathbb{R}^d$. Suppose that $(g_0, \theta_0) = \arg\min_{g, \theta}\text{KL}(P_0, P_{g, \theta})$ exists and is unique. Let $\mu + \sigma P_{Z \mid \theta, X}$ be identifiable with respect to $(\mu, \sigma, \theta)$ and assume prior independence
$\Pi_{\mu,\sigma,\theta} = \Pi_{\mu,\sigma} \times \Pi_\theta$.  Suppose that $p(\tilde g \mid \D_n)$ is (strongly) consistent at  $\mu_0 + \sigma_0 \tilde g_0$, where $\mu_0$ and $\sigma_0 \ne 0$ are constants and $(\mu_0, \sigma_0) \in \text{supp}(\Pi_{\mu, \sigma})$.  Under the key assumption
\begin{enumerate}[(\ref{thm:theta-robustness}.1)]
\vspace{-2mm}\item there exists a neighborhood $\mathcal{G}$ around $\mu_0 + \sigma_0 g_0$ under the topology $\mathcal{T}$ such that for any $g_n \to g$ in $\mathcal{G}$, $\sup_{\mu \in \R, \sigma \in \R^+, \theta \in \Theta}|KL(P_0, P_{(g_n - \mu)/\sigma, \theta}) - KL(P_0, P_{(g - \mu)/\sigma, \theta})| \to 0$, 
\end{enumerate}
and if assumptions (\ref{thm:theta-con}.1)--(\ref{thm:theta-con}.4) hold for all $g$, then the joint posterior distribution $p(\tilde g, \theta \mid \D_n)$ is (strongly) consistent at $(\mu_0 + \sigma_0 \tilde g_0, \theta_0)$. 

\end{theorem}
This theorem provides robustness guarantees for the model \eqref{trans}--\eqref{mod} under double misspecifications of both $\theta$ and $g$, and in particular ensures marginal (strong) consistency of $p(\theta \mid \D_n)$ at $\theta_0$. Specifically, $\theta$ may be misspecified in the sense that the true parameter is not contained in the parameter set $\Theta$, and $g$ may be misspecified as a location-scale shift around $g_0$, where $g_0$ is a fixed monotonic transformation. Notably, this result does not require $g_0$ and $\theta_0$ to be the true parameters, but instead establishes posterior consistency for any pair of the transformation $g_0$ and the KL-minimizer $\theta_0$, as long as the conditions are satisfied. The main condition,   (\ref{thm:theta-robustness}.1) appears complex but is a common requirement in the asymptotic analysis of misspecified Bayesian semiparametric models. 
It is a variant of similar conditions related to posterior convergence under perturbations around ``least-favorable submodels" of the true model \citep{Bickel2012}.

\section{Discussion} \label{sec-disc}
We introduced a Bayesian approach for semiparametric regression analysis. Our strategy featured a transformation layer \eqref{trans} atop a continuous regression model \eqref{mod} to enhance  modeling flexibility, especially for irregular marginal distributions and various data domains. Most uniquely, the proposed sampling algorithm (Algorithm~\ref{alg:joint}) delivered efficient Monte Carlo (not MCMC) inference with easy implementations for popular regression models such as  linear regression, quantile regression, and Gaussian processes. Empirical results demonstrated exceptional prediction, selection, and estimation capabilities  for Bayesian semiparametric linear models; substantially more accurate quantile estimates and model adequacy for Bayesian quantile regression; and superior predictive accuracy for Gaussian processes. 
Finally, our asymptotic  analysis established joint posterior consistency under general conditions, including multiple model misspecifications.

The primary concerns with Algorithm~\ref{alg:joint} are (i) the use of the surrogate likelihood in place of \eqref{exact-like}  and (ii) the need for an approximation $\hat p(\theta \mid \D_n)$ to infer  $g$ via \eqref{Fzx}. We have attempted to address these concerns using both empirical and theoretical analysis. The empirical results are highly encouraging: the  Monte Carlo samplers are efficient and simple for a variety of semiparametric Bayesian models and the posterior predictive distributions are calibrated and sharp for many challenging  simulated and real datasets. These results  are robust to the choice of the approximation, including simply the prior $\hat p(\theta \mid \D_n)  = p(\theta)$. Further, we introduced an importance sampling adjustment to provide posterior inference under the correct  likelihood. Yet this adjustment does not appear to make any difference in practice, even for $n=50$, which is reassuring for direct and unadjusted application of Algorithm~\ref{alg:joint}. 
Finally, the theoretical analysis establishes the asymptotic validity of the  posterior targeted by Algorithm~\ref{alg:joint},  even under multiple model misspecifications. This remains true for a broad class of (consistent) marginal models for $P_Y$ and $P_X$, which allows customization for settings in which the Bayesian bootstrap is not ideal. 

\ifblinded

\else 
\section*{Acknowledgements}
We thank the associate editor, the associate editor of reproducibility, and two anonymous referees for their constructive comments that substantially improved the quality of the article.  We also thank David Ruppert and Surya Tokdar for their helpful comments. 
Research (Kowal) was sponsored by the Army Research Office (W911NF-20-1-0184), the National Institute of
Environmental Health Sciences of the National Institutes of Health (R01ES028819), and the National Science
Foundation (SES-2214726). The content, views, and conclusions contained in this document are those of the
authors and should not be interpreted as representing the official policies, either expressed or implied, of the
Army Research Office, the National Institutes of Health, or the U.S. Government. The U.S. Government
is authorized to reproduce and distribute reprints for Government purposes notwithstanding any copyright
notation herein.
\fi

\setstretch{1.5}
\bibliographystyle{apalike}
\bibliography{refs.bib}
\end{document}